\documentstyle[12pt,aasms]{article}

\include{epsf}

\begin{document}

\newcommand{\be}{\begin{equation}}
\newcommand{\ee}{\end{equation}}
\newcommand{\ba}{\begin{eqnarray}}
\newcommand{\ea}{\end{eqnarray}}
\newcommand{\brr}{\begin{array}}
\newcommand{\err}{\end{array}}
\newcommand{\bc}{\begin{center}}
\newcommand{\ec}{\end{center}}
\newcommand{\br}{\mbox{\bf r}}
\newcommand{\bx}{\mbox{\bf x}}
\newcommand{\by}{\mbox{\bf y}}
\newcommand{\bk}{\mbox{\bf k}}
\newcommand{\lb}{{\left<\right.}}
\newcommand{\rb}{{\left.\right>}}
\newcommand{\eps}{\epsilon}
\newcommand{\barG}{\bar\Gamma}
\newcommand{\G}{\Gamma}
\newcommand{\bu}{{\bf u}}
\newcommand{\bW}{{\bf W}}
\newcommand{\D}{\Delta}

\def\cenlimpv/{1}
\def\denstate/{2}
\def\vrscatt/{3}
\def\longpdf/{4}
\def\autocorr/{5}
\def\traj/{6}
\def\vtfast/{7}
\def\nnfast/{8}
\def\distratio/{9}
\def\vordisp/{10}
\def\drboth/{11}
\def\diffusion/{12}

{\baselineskip=18truept

\title{Lagrangian Dynamics in High-Dimensional Point-Vortex Systems}

\centerline{Jeffrey B. Weiss,\footnote[1]{
PAOS, Campus Box 311, University of Colorado, Boulder, CO 80309-0311,
USA, \vskip -10truept jweiss@colorado.edu.
}  
Antonello Provenzale,\footnote[2]{
JILA, Campus Box 440, University of Colorado, Boulder, CO 80309-0440,
USA; \vskip -10truept permanent address: Istituto di Cosmogeofisica
del CNR, C.so Fiume 4, I-10133 Torino, \vskip -10truept Italy.
}
and James C. McWilliams\footnote[3]{IGPP, UCLA, Los Angeles, CA
90095-1567.}
}  

\medskip

\centerline{\bf Abstract}

We study the Lagrangian dynamics of systems of $N$ point vortices and
passive particles in a two-dimensional, doubly periodic domain.  The
probability distribution function of vortex velocity, $p_N$, has a
slow-velocity Gaussian component and a significant high-velocity tail
caused by close vortex pairs.  In the limit for $N\to\infty$, $p_N$
tends to a Gaussian.  However, the form of the single-vortex velocity
causes very slow convergence with $N$; for $N\approx 10^6$ the
non-Gaussian high-velocity tails still play a significant role.  At
finite $N$, the Gaussian component is well modeled by an
Ornstein-Uhlenbeck (OU) stochastic process with variance $\sigma_N =
\sqrt{N \ln N/2\pi}$.  Considering in detail the case $N=100$, we show
that at short times the velocity autocorrelation is dominated by the
Gaussian component and displays an exponential decay with a short
Lagrangian decorrelation time.  The close pairs have a long
correlation time and cause nonergodicity over at least the time of the
integration.  Due to close vortex dipoles the absolute dispersion
differs significantly from the OU prediction, and shows evidence of
long-time anomalous dispersion.  We discuss the mathematical form of a
new stochastic model for the Lagrangian dynamics, consisting of an OU
model combined with long-lived close same-sign vortices engaged in
rapid rotation and long-lived close dipoles engaged in ballistic
motion.  From a dynamical-systems perspective this work indicates that
systems of dimension $O(100)$ can have behavior which is a combination
of both low-dimensional behavior, i.e. close pairs, and extremely
high-dimensional behavior described by traditional stochastic
processes.

}

\centerline{PACS: 47.32.Cc, 47.52.+j, 47.27.Qb}

\centerline{January 22, 1998}
\centerline{to appear in {\it Physics of Fluids}}

\vfill\eject

\renewcommand{\thefootnote}{\fnsymbol{footnote}}

\section{Introduction}

Advection of passive and active tracers is a crucial component in many
geophysical processes: ozone transport in the stratosphere,
pollutant dispersal in the atmosphere and the ocean, plankton
and salinity transport in the ocean. Additionally,
understanding the relationships between Eulerian and Lagrangian
statistics is necessary for interpreting data provided by ocean floats
and atmospheric balloons.

Standard Lagrangian approaches to transport in fluids are based on the
use of either stochastic models or chaotic advection in
low-dimensional dynamical systems. Lagrangian stochastic models are
generally based on assuming that transport is dominated by
unstructured Gaussian random velocity fluctuations, and they are most
successful when the system under study has an extremely
high-dimensional phase space.$^{\cite{Adler,Wax}}$ The standard
example is Brownian motion which describes the irregular movement of
microscopic particles in systems with dimensionality of the order of
Avogadro's number, $O(10^{23})$. In this case, the details of the
deterministic description are irrelevant.

Chaotic advection, on the other hand, is based on a fully
deterministic description of the phase-space dynamics. It has
traditionally been applied to systems with just a few excited degrees
of freedom, with very simple spatial structures, and with periodic or
quasi-periodic temporal dynamics.$^{\cite{Ottino,PhysicaD,CSF}}$ Thus,
the traditional application of stochastic models is to highly
turbulent flows without strong coherent structures, while
low-dimensional chaotic advection is appropriate for flows dominated
by a few large-scale waves, vortices, or other
structures.$^{\cite{Weiss91,Weiss94}}$

Geophysical turbulence, however, does not fully belong to either of
the above categories.  One of the main difficulties encountered in
studying Lagrangian advection in geophysical turbulence is, in fact,
the complex space-time structure of the flow.  The joint effects of
rotation and stratification often induce the presence of energetic
coherent structures that contain the majority of the enstrophy of the
system.$^{\cite{MWY94,MW94,ML91}}$ Intense jets such as the Gulf
Stream act as partial barriers to
transport, while coherent vortices such as ocean mesoscale eddies and
the stratospheric polar vortex can trap particles for long
times.$^{\cite{SRC95,EPB93,Samelson,M85,Papa97}}$ Previous studies of
barotropic turbulence have shown that coherent vortices play an
important role in the advection of Lagrangian tracers, inducing
characteristic signatures that cannot be captured by simple stochastic
models.$^{\cite{EPB93,Babiano94,Provenzale97}}$

In this work, we further explore the properties of advection in flows
dominated by strong coherent vortices, and, in particular, we consider
an ensemble of many point vortices in two spatial dimensions.  Systems
of points vortices capture many of the features of two-dimensional
turbulent flows.$^{\cite{Weiss98,Riccardi95,WM93,BCBS92,Carn91}}$ In
geophysically relevant situations the number of energetic coherent
structures is neither as large as Avogadro's number, nor as small as
the number of degrees of freedom in simple models of low-dimensional
chaotic advection.  Systems of $N$ point vortices with $N \sim
O(10^m)$, $m$ small, are thus used here as another simplified
paradigm: they have an intermediate phase-space dimensionality, and
are appropriate for describing advection in vortex dominated flows.
As we show below, the dynamics of systems of many point vortices
displays properties of both stochastic models and low-dimensional
chaotic advection, and provides a useful bridge between
high-dimensional stochastic models and low-dimensional chaotic
advection.

The remainder of this paper is organized as follows. Section 2 contains
an introduction to the dynamics of point vortices, Section 3 addresses
the central limit theorem for point-vortex systems and
considers the statistical properties of ensembles of point
vortices. In Section 4 we discuss the statistical properties of long
time-integrations of systems of 100 point vortices, and study the
behavior of both the vortices and of passively advected particles. In
section 5 we study single-particle dispersion in point-vortex systems
and compare with the dispersion of an ensemble of randomly moving
particles described by the Ornstein-Uhlenbeck (OU) stochastic
process.$^{\cite{Wax}}$  Finally, in Section 6 we present conclusions
and perspectives, and discuss a possible alternative stochastic model
for describing advection in point-vortex systems.

\section{Point-Vortex Dynamics}

Point vortices are singular solutions of the Euler equations in
two spatial dimensions. The dynamics of an ensemble of point vortices 
is described self-consistently by a system of equations that take the
non-canonical Hamiltonian form
\be
\G_i{{dx_i}\over {dt}} = {{\partial H}\over{\partial y_i}},
\qquad
\G_i{{dy_i}\over {dt}} = -{{\partial H}\over{\partial x_i}},
\label{pv}
\ee
where $\bx_i = (x_i,y_i)$ is the position of the $i$-th vortex with
constant circulation $\G_i$, $H$ is the Hamiltonian
\be
H\left(\left\{\bx_i\right\}\right) =  
-\sum_{{i,j=1 }\atop {i\ne j}}^N {{\G_i \G_j}\over {2}}  
G(\bx_i,\bx_j),
\label{Hgen}
\ee 
$N$ is the number of vortices, and the form of the Green function $G$
depends on the boundary conditions.$^{\cite{Aref}}$ The positions $x_i$
and $y_i$ play the role of non-canonically conjugate variables, the
number of degrees of freedom is equal to the number of vortices, and
the dimensionality of the phase space is twice the number of vortices.
The number of independent conserved quantities of (\ref{pv}) and
(\ref{Hgen}) also depends on the boundary conditions, but it is always
finite and small. Hence, an assembly of more than a few vortices
behaves chaotically.   

Passively advected particles are easily incorporated in point-vortex
systems.  A point vortex with zero circulation, $\G_i = 0$, is
advected by the other vortices but has no influence on the velocity of
any other vortex.  Such passively advected particles will be referred
to here as {\it passives}, and particles with $\G_i \ne 0$ as
vortices.  Recently, the motion of passive particles in point-vortex
systems has been the subject of several
investigations.$^{\cite{Babiano94,Boffetta96,Tel95}}$ These studies have
shown that point vortices are surrounded by finite-size islands of
regular Lagrangian motion, where passives can be trapped for very long
times. This implies the existence of long-time non-ergodicity in
passive particle motion, where averages over ensembles of different
advected particles differ from time averages over single-particle
trajectories. In turn, this may lead to long lasting differences
between Eulerian and Lagrangian averages and the possibility of
non-Brownian (anomalous) dispersion. In the following, we explore in
detail some of these issues.

In this work we use periodic boundary conditions because it is the
only domain that has the needed properties.  A closed
domain has a finite maximum separation and is thus not suited to study
long-time dispersion properties.  This leaves the infinite plane, the
periodic plane (2-torus), and the sphere.  The infinite plane is
unsuitable because motion is not homogeneous with a finite number of
vortices.  On a sphere, there is no unique way to count the number of
times a particle travels around the domain, and thus the sphere does
not allow a satisfactory definition of particle dispersion at long
times.  Thus the only boundary condition which meets our requirements
is that of the periodic domain.  In addition, most simulations of
homogeneous turbulence use these boundary conditions.

The Green function $G$ for point vortices on a periodic domain with
length $2\pi$ may be written as$^{\cite{Campbell,WM91}}$
\be
G(\bx_i,\bx_j) = \sum_{m=-\infty}^{\infty} 
\ln\left({\cosh\left(x_i - x_j -2\pi m\right) - \cos\left(y_i - y_j\right) 
\over  \cosh\left(2\pi m\right)}\right) 
- {\left(x_i - x_j\right)^2 \over 2\pi}.
\label{ham}
\ee
The function $G$ can be shown to be periodic in $x$ and $y$, and
invariant under the transformation $x \rightleftharpoons y$.  The
velocity of the $i$-th vortex resulting from (\ref{pv}), (\ref{Hgen}),
and (\ref{ham}) is a sum over the velocities induced by each of the
other vortices:
\vfill\eject
\ba
u_i =\sum_{{j = 1}\atop{j\ne i}}^N \G_j 
\sum_{m=-\infty}^{\infty}
	{{-\sin(y_i - y_j)}\over{\cosh(x_i-x_j-2\pi m) - \cos(y_i - y_j)}},
\nonumber\\
v_i=\sum_{{j = 1}\atop{j\ne i}}^N \G_j 
\sum_{m=-\infty}^{\infty}
	{{\sin(x_i - x_j)}\over{\cosh(y_i-y_j-2\pi m) 
	- \cos(x_i - x_j)}},\label{vel} 
\ea
where $\bu = (u_i,v_i) = (dx_i/dt, dy_i/dt)$.  The dynamics on the periodic
domain has two invariants, the components of the linear momentum,
corresponding to translation parallel to the $x$ and $y$ axes;
$P_x=\sum_{i=1}^N \G_i y_i$ and $P_y=\sum_{i=1}^N \G_i x_i$. Angular
momentum is not an invariant because the periodic boundary conditions
break the rotational symmetry.  The invariants $P_x$ and $P_y$ are
independent only if the total circulation $\sum \G_i$ is
zero.$^{\cite{Aref}}$ In this case, the motion is chaotic for $N > 3$.

\section{Point-Vortex Configurations}

In this section we discuss the properties of instantaneous
configurations of $N$ vortices. We restrict ourselves to
configurations with zero total circulation and an equal number of
positive and negative vortices.

First consider the velocity induced at a point $\bx_0 = (x_0, y_0)$ by
a single vortex at $\bx_1$.  In the limit $r = |\bx_1 - \bx_0| \to 0$
the velocity (\ref{vel}) asymptotically approaches the velocity
generated by a single vortex on the infinite domain, $2
\G/r$.\footnote{Note that with the choice (\ref{ham}) for the Green
function, the timescale differs from that usually used on the infinite
plane by a factor of $4\pi$; hence the velocity from a close vortex is
here a factor $4\pi$ greater than the expression usually used
on the infinite domain.$^{\cite{Aref}}$} At larger distances
from the vortex, the domain periodicity modifies this simple form of
the velocity.  In practice, the velocity on the periodic domain
departs significantly from the infinite plane result only for
separations greater than approximately one sixth of the domain.

Next, consider the probability density function (pdf) for the velocity
resulting from a single vortex, $p_1(\bu)$.  This has been studied in
detail by Min et al in the case of the infinite
domain.$^{\cite{Min96}}$ If the position of the vortex is chosen at
random from a uniform distribution, then the high velocity limit of
$p_1$ is easily calculated.  Since the probability of having
separation between $r$ and $r+dr$ is $2\pi r dr$, and since for small
$r$ the velocity scales as $1/r$ then for large $|\bu|$, $p_1(\bu)
\sim p(r(|\bu|)) \left | dr/d|\bu| \right| \sim 1/|\bu|^3$.  This
scaling for large velocities is independent of the boundary
conditions, as the boundaries are irrelevant at sufficiently small
separations.  The detailed form of the distribution for smaller
$|\bu|$ does, however, depend on the specific form of $\bu$ for a
periodic domain, and it depends on the direction of $\bu$ as well as
its magnitude.

Since we are interested in the dynamics of systems of several point
vortices, we now consider the velocity induced at a point by a system
of $N$ vortices with $|\G_i| = 1$.  This velocity 
is merely the sum of the velocities due to each vortex separately.  If
the vortices are randomly placed, then the velocity is the sum of $N$
random numbers, and its pdf $p_N(\bu)$ is determined by $p_1(\bu)$.
For functions $p_1$ which decay sufficiently rapidly, the central
limit theorem applies and $p_N$ approaches a Gaussian.  Most
statements of the central limit theorem require that the variance of
$p_1$ be finite.  Here, since $p_1 \sim 1/|\bu|^3$ the variance $\int
|\bu|^2 p_1(|\bu|) d |\bu|$ diverges logarithmically.  However, this
divergence is sufficiently slow that the central limit theorem still
applies and as $N \to \infty$, $p_N$ does become Gaussian.  For more
details, see the Appendix, Min et al$^{\cite{Min96}}$, and Ibragimov and
Linnik.$^{\cite{Ibrag}}$

For large but not infinite $N$, the velocity pdf of an ensemble of
vortices displays interesting properties.  For distributions with
finite variance, the convergence to a Gaussian is quite rapid.  For
example, if $p_1$ were constant over a finite region and zero
elsewhere, then $p_N$ is quite close to Gaussian even for $N=5$.
However, for distributions with slowly diverging variance, such as the
case here, the convergence to a Gaussian is slow and is called
``non-normal''.$^{\cite{Ibrag}}$ For non-normal convergence, the
variance of the sum scales differently than the normal case.  In the
Appendix we show that for point vortices the variance of the
asymptotic Gaussian is $\sigma_N = \sqrt{N \ln N/ 2 \pi}$.
Figure~\cenlimpv/ shows how $p_N(u/\sigma_N)$ asymptotically
approaches a normal distribution.  Due to symmetry, $p_N(u) = p_N(v)$
and these two distributions have been combined into a single pdf.
Further, $p_N$ is even and we combine positive and negative
velocities.  A least-squares fit of the small velocity portion of the
distribution to a Gaussian provides a variance of approximately one
(as expected for $N\to \infty$) and an amplitude $0.9$ of that of a
normal distribution.  One sees that even though the
central limit theorem formally applies, the convergence is extremely
slow and even for large $N$, $p_N$ has significant high-velocity
tails.  Further, the small velocity part of $p_N$ is well approximated
by a Gaussian with the asymptotic variance, but with amplitude less
than one due to the significant fraction of events contained in the
high-velocity tails.  In what follows, we show that both the Gaussian
portion of the pdf at small velocities and the high-velocity tails
play an important role in the advection process.

In the remainder of this paper we focus on systems of $100$ point
vortices. Similar results are obtained with other values of $N$
between about 10 and a thousand.  From the perspective of dynamical
systems, an ensemble of 100 point vortices has a $200$-dimensional
phase space; it is thus quite high dimensional. However, it is still
reasonable to numerically integrate the equations for significant
times.  From the perspective of the central limit theorem, due to the
slow non-normal convergence, $N=100$ is not large enough for $p_{100}$
to be completely Gaussian.

We next consider several different initial configurations of the $N$
vortices.  The circulations of the vortex populations are
characterized by $\barG$, the average of the absolute value of the
individual circulations.  We focus on configurations with $\barG =
1$, and with random individual circulations uniformly
distributed in the range $0.8\barG \le |\G_i| \le 1.2\barG$.  The
variation in the individual $\G_i$'s is included to break any symmetry
in vortex pair interactions that may arise in a system with identical
$|\G_i|$'s.  For each configuration, the initial vortex positions are
randomly chosen in the square domain $[0,2\pi]^2$.  For simplicity we
only consider initial conditions with $P_x=P_y=0$.

The set of initial configurations discussed above can have a range of
energies, $E = H\left(\left\{\bx_i\right\}\right)/4 \pi^2$. Close
opposite-sign vortex pairs give a large negative contribution to the
energy, while close same-sign pairs provide a positive
contribution. Thus, depending on the random initial positions of the
vortices, the energy of the system may take different values. In
Figure~\denstate/ we show the density of states, i.e., the number of
states with energy in a given interval, obtained from $10^4$ different
realizations of the initial vortex configurations described above with
$N=100$.  This distribution has a mean of -0.076, a variance of 1.18,
and a skewness coefficient of 0.85.  In the following we integrate the
motion of 300 random configurations extracted from this ensemble for a
relatively short time, $T = 0.1$. The two vertical dotted lines
indicate two arbitrarily chosen configurations that have been
integrated for a much longer time, $T = 30$.

The relationship between vortex speed and nearest neighbor distance
for all the $100$ vortices in the $300$ randomly chosen configurations
is seen in Figure~\vrscatt/.  Large speeds are associated with close
nearest neighbors.  From this we may conclude that the high velocities
in the tail of the pdfs are due to the presence of a single close
vortex and not to the superposition of the contributions of several
different vortices. Thus, high-velocity tails in $p_{100}$ are
essentially two-vortex phenomena and not many-body effects. This
implies that in the limit $N\to \infty$, as the tails of $p_N$
disappear, the importance of two-vortex interactions goes to zero.
However, given the slow convergence of $p_N$, two-vortex interactions
will be important even for very large $N$.

\section{Long-Time Integrations}

In this section we focus on the phase-space trajectories obtained by a
relatively long-time integration of the two randomly selected initial
configurations indicated in Figure~\denstate/.  These initial
configurations have $N=100$ vortices and $N_p=100$ passives. 
The configurations were integrated until $T=30$ using a
fourth-order Runge-Kutta method with fixed time step $\Delta t =
10^{-5}$. The positions and velocities of the vortices were saved
every $\Delta t_s = 10^{-2}$.  

The Cartesian velocity pdfs, averaged over time and all vortices,
denoted by $\bar p(u)$, are obtained by measuring $u$ and $v$ for all
vortices every $\Delta t_s$ throughout the integration; again, from
symmetry $\bar p(u) = \bar p(v) = \bar p(-u) = \bar p(-v)$ so $\pm u$
and $\pm v$ are combined into a single pdf.  The pdfs for the vortex
and passive velocities in each of the two solutions are shown in
Figure~\longpdf/.  The central part of each pdf, $|u| \lesssim 20$, is
well approximated by the same Gaussian which fits the central part of
$p_{100}$; it has the theoretical asymptotic width $\sigma_{100} =
\sqrt{100 \ln 100/ 2 \pi} \approx 8.56$ and amplitude $0.9$ of a
Gaussian with unit normalization.  Thus the vortices and passives in
the two solutions all have the same small-velocity pdf as each other
and as an ensemble of random initial conditions.

The high-velocity tails of $\bar p$ for the two solutions are,
however, significantly different.  One solution has a significant
excess over $p_{100}$ at large velocities, while the other has a
deficit.  Note that both solutions, however, have a deficit at
extremely large velocities, $u \gtrsim 250 \approx 30 \sigma_{100}$.  
The pdf with smaller tail in the vortices comes from the same
integration as the pdf with the larger tail in the passives, and
vice-versa.  This is not significant, and is merely the result of
picking only two random initial conditions.  

The lack of convergence of the pdfs over the time of the integration
$T$ indicates that $T$ is not long enough for the system to forget its
initial conditions. By other measures, however, $T$ is very long.  A
typical eddy turnover time, estimated as the time for a vortex at a
typical vortex separation $2\pi/\sqrt{N}$ from another vortex to
rotate in a complete circle, is approximately $0.4$. Another measure
of the eddy turnover time is $\left( \sum \G_i^2 \right)^{1/2} \approx
0.1$. Thus $T$ is $O(100)$ eddy turnover times.  A completely
different measure of the length of the integration is that in a time
of $T$ the particles travel around the basic periodic domain several
times.  All of this indicates that the correlation time for the tail
of the pdf is very long, and that the point-vortex system displays a
very long memory with an associated lack of ergodicity over large, if
not infinite, time intervals. While this is consistent with what has
already been observed for systems of a few point
vortices,$^{\cite{Babiano94,WM91}}$ it is somewhat surprising that
this phenomenon persists in a system with so many degrees of freedom.

A standard statistic for studying velocity time series is the
Lagrangian velocity autocorrelation $R(\tau)$,
\be
R(\tau) = {{\lb {\bf u}(t) \cdot {\bf u}(t +\tau)\rb} 
\over {\lb |{\bf u}(t)^2|\rb} },
\label{autodef}
\ee where $\lb \cdots\rb$ represents an average over time and over
particles.  The autocorrelations from the long phase-space
trajectories for the 
two populations of vortices and passives are shown in Figure~\autocorr/.
As in the pdfs, there is no significant difference between the
vortices and passives.  After a very short period with steep decay,
the autocorrelation of the vortices and passives follows an
exponential extremely well. For $\tau$ greater than those shown in the
Figure, $R(\tau)$ oscillates around zero.  The Lagrangian
decorrelation time of the phase-space trajectories, $T_L$, is
estimated by fitting 
$R(\tau) \sim \exp(-\tau/T_L)$ for the period $\tau=0.02$ to $\tau =
0.2$, giving $T_L = 0.09 \pm 0.01$.  A theoretical estimate of the order of
magnitude of $T_L$ can easily be obtained by assuming that $T_L$ is
the time it takes for a vortex moving at a typical velocity $\sigma_N
= \sqrt{N \ln N/ 2 \pi}$ to cross a distance equal to the typical
vortex separation $2 \pi/\sqrt{N}$; this estimate gives $T_L \sim (2
\pi)^{3/2}/N \sqrt{\ln N} \sim 0.07$, quite close to that
effectively observed. Note also that the common method of estimating
the Lagrangian integral time scale, $T_I=\int_0^\infty R(\tau) d\tau$,
provides misleading results because the very long memory of
the system implies that extremely long integration times are needed to
reliably calculate $T_I$.

Individual vortex and passive trajectories, some
examples of which are shown in Figure~\traj/, are in general quite
complex.  The first two panels of Figure~\traj/ show examples of an
intermittently fast moving vortex and of a vortex randomly selected
from the population of vortices with moderate average speed.  The
single fast vortex shown in Figure~\traj/(a) comprises 33\% of the
extreme tail (velocity greater than $10 \sigma_{100})$ of the
velocity pdf $\bar p$ for its integration.  Figure~\traj/(c) shows the
particle trajectory of a randomly selected passive particle.  The
particle trajectories of the slow vortex and the passive are
qualitatively similar, 
although the passive does have more tight loops, indicating that at
those times it is close to a vortex.  The particle trajectory of the
fast vortex is qualitatively different than the other two.

The time series of the speed of the fast vortex, Figure~\vtfast/,
shows clear bimodal behavior with the vortex jumping between slow and
fast episodes.  A time series of the distance to the closest vortex
and its identity, Figure~\nnfast/, shows that episodes of extremely
fast motion coincide with the vortex being close to another,
oppositely-signed, vortex, i.e., it is a member of a close dipole.
These episodes start and end with changes in the identity of the close
vortex.  Furthermore, the three separate fast episodes are separated
by extremely short periods of slow motion where the closest vortex is
significantly further away and changes identity rapidly.  The ratio of
the distance of the second closest vortex to the distance of the
closest vortex, Figure~\distratio/, demonstrates that the fast
episodes are primarily the result of close dipoles.  However, note
that during the period from $t\approx 22$ to $t \approx 24$ this
distance ratio is smaller than during other fast periods (but still
larger than during slow periods), and there are small oscillations in
the nearest neighbor distance (Figure~\nnfast/).  Analysis of the
individual vortex trajectories shows that during this period the
second closest vortex has the same identity and there is thus a
temporary three-vortex bound state.

Before closing this section, we note that we have not detected any
significant difference between vortex and passive statistics, in the
sense that the internal variability in each of these populations is by
far larger than any difference between vortices and passives. This is
not too surprising if we consider that the contribution of each individual
vortex to the total vorticity field is rather small for $N=100$, and
each individual vortex in the collective field of the other vortices
behaves similar to a passive.  Significant differences do exist,
however, in strong two-body interactions: close vortices accelerate
each other and may undergo episodes of fast displacement, 
while passives trapped in the vicinity of a vortex do
not influence its motion.   

When passives are close to a vortex, they spin with large velocity and
create high-velocity components in the pdf.  Moreover, passives close
to a vortex tend to remain associated with it for very long times,
both on the infinite plane$^{\cite{Babiano94}}$ and in periodic
domains.$^{\cite{Care}}$ The presence or absence of high-velocity
"bumps" in the pdf is thus entirely determined by the initial
conditions, i.e., whether or not the system happens to start with
close passives.  Only by integrating for times much longer than the
trapping time is it possible for the time average pdf to converge to
the ensemble average pdf, an issue which is related to the possible
non-ergodicity of passive particle motion in point vortex
systems.$^{\cite{Babiano94}}$ At present, it is unclear whether the
trapping islands around the vortices have a time-asymptotic nature or
disappear at finite (but long) time.  Preliminary runs of few-vortex
systems$^{\cite{Care}}$ have shown that the trapping islands exist up
to at least $t=3000$. By comparison, bound states of vortex couples
tend to live for much shorter times.

When a passive is close to a vortex, it thus behaves similar to the
vortex with which it is linked, with an additional fast rotational
component.  Whenever the vortex undergoes a fast displacement due to
vortex dipolar coupling, the trapped passive does as well. A detailed
theoretical comparison of the long-time displacement statistics of
vortices and passives should thus carefully evaluate the probability
that a passive be trapped near a vortex (which is related to the
initial conditions and determined mainly by geometric factors) and the
lifetime of vortex-vortex and vortex-passive couples.  On a purely
heuristic basis, the present simulations have shown no detectable
differences between the gross statistical behavior of vortices and
passives, at least on the time scales we have considered.

\section{Properties of single-particle dispersion}

In this section we study the properties of single-particle dispersion
in point-vortex systems.  A classic stochastic model for describing
the motion of advected particles is the Ornstein-Uhlenbeck (OU)
process$^{\cite{Wax,vanDop}}$
\be
d{\bf x}={\bf u}\,dt, 
\qquad
d{\bf u} = -{{\bf u}\over{T_L}} dt
+ {\sigma_{ou}\over{T_L^{1/2}}} d{\bf W},
\label{ou}
\ee 
where $\bW$ is the Wiener process, and $d\bW$ is a Gaussian
random increment with $\lb d\bW \rb=0$ and $\lb
dW_\alpha(t)\,dW_\beta(t')\rb =2\,\delta_{\alpha,\beta}\,
\delta(t-t')\,dt$, where $\lb \cdots\rb $ indicates an average over an
ensemble of independent realizations, and Greek subscripts indicate
vector components.  The OU process is characterized by two parameters:
a velocity scale $\sigma_{ou}$ which determines the variance of the
velocity pdf, and a timescale $T_L$ which is the Lagrangian
decorrelation time of the exponential velocity autocorrelation 
$R_{ou}(\tau) = \exp(-\tau/T_L)$. Recently, the OU process
has been used to model single-particle dispersion in the atmosphere
and the ocean.$^{\cite{Griffa}}$

Here, the OU process is a very natural choice for a stochastic model.
As we have already seen, the velocity pdf has a significant Gaussian
component, leading us to pick a Gaussian process.  Additionally, the
Lagrangian velocity autocorrelation has a period of exponential decay,
which leads to the OU process.  Furthermore, the two parameters of the
OU model have already been determined: $\sigma_{ou} =
\sigma_{100}\approx 8.56$ and $T_L = 0.09$.  The major failure of the
OU process is that it does not capture the high-velocity tails of the
vortex pdfs.  A trajectory of an OU particle with
these parameters is shown in Figure~\traj/(d). The large scale motion is
similar to that of a slow vortex or passive, but the small scale
motion has more sharp turns instead of small loops.

The OU process may thus give a framework for analyzing the time
evolution of the vortex statistics, at least for the slow portion of
the vortex population.  This is in contrast to the application of the
central limit theorem in Section 3, which only refers to single
instants in time.  To explore similarities and differences with the OU
process, relatively short-time integrations, until $T = 0.1$, were
performed on $300$ initial conditions chosen randomly from those shown
in Figure~\denstate/.  Each initial condition has $100$ vortices and $100$
passives.  This dataset is thus the same size as a single long
integration.  The velocity pdfs averaged over all initial conditions,
as well as the time and initial condition averaged pdfs, are equal to
$p_{100}$. 

We now compare the vortex displacements, $\D x(\tau) = x(\tau)- x(0)$
with the displacements of the OU process.  As is usually done in the
study of systems with periodic boundary conditions, we compute
displacements from the unfolded trajectories, i.e. by taking into
account the number of times a particle has wound around the periodic
domain.$^{\cite{EPB93}}$ Again, due to symmetry, we can combine $x$
and $y$ displacements.  The OU process has displacements with a
Gaussian pdf whose width $\sigma_{\D x}$ is 
\be 
\sigma^2_{\D x}(\tau)
= 2 \sigma^2_{ou} T_L^2 \left( {{\tau}\over{T_L}} + e^{-\tau/T_L} - 1
\right).
\label{Dxou}
\ee 
The pdf of vortex displacements at
time delay $\tau$ can be compared to the OU pdf by scaling each
vortex displacement by the above $\sigma_{\D x}(\tau)$.  The resulting
set of pdfs, as obtained by considering all 100 vortices for all 300
initial configurations, are shown in Figure~\vordisp/ for time delays
$\tau=0.01,0.02,...,0.09$. This figure shows that for small
displacements the pdfs agree with the OU result.  However, there are
significant large-displacement tails, which are largest for small times,
and decrease as $\tau$ increases.

To identify the origin of the large-displacement tails, in
Figure~\drboth/ we show the displacements at $\tau = 0.05$ versus the
nearest neighbor distance, for both same-sign and opposite-sign
neighbors.  The large displacements are associated with close
opposite-sign vortices, i.e. a vortex dipole, while there is no
correlation between large displacements and close same-sign
vortices. Further, the value of the large displacement is well
predicted by the motion of an isolated dipole.  

We can conclude that the small displacements are essentially a
mean-field process, in that 1) they are due to the action of many
other vortices and 2) they are well captured by a stochastic model
such as the OU process, which is known to work well for extremely
high-dimensional systems such as Brownian motion.  On the other hand,
the large displacements are essentially due to two-body interactions,
namely, temporary coupling of opposite-sign vortices.  Same-sign
couples rotate rapidly about their center of vorticity, and otherwise
behave similarly to a single vortex with larger circulation.  Thus,
they contribute to the high-velocity tails, but do not contribute to
the large-displacement tails.  Similar results are observed for the
passive particles, with the usual caveat that when passives couple
with a vortex they do not affect its velocity, but they can be trapped
in the vicinity of a vortex undergoing fast dipolar motion.

We next turn to the temporal evolution of the average vortex and
passive displacement, i.e, to absolute or single-particle dispersion.
The single-particle dispersion of a system of moving particles is
defined as 
\be
A^2(\tau,t) = \lb  \left(\bx(t+\tau) - \bx(t)\right)^2\rb_p,
\label{dispdef}
\ee
where $\lb \cdots\rb_p$ indicates an average over the ensemble of
particles.  When the particle motion is statistically stationary, as
is the case here, $A^2$ becomes independent of $t$.  The finite-time
dispersion coefficient is defined as $D(\tau)= A^2 / 2 \tau$.  For
motions in differentiable velocity fields, the short time dispersion
is ballistic, i.e. $A^2(\tau) \propto \tau^2$ for $\tau << T_I$ where
$T_I$ is the Lagrangian integral time scale, defined as the integral
of the velocity autocorrelation.  If $T_I$ is finite, then at large
times the dispersion becomes Brownian-like, i.e. $A^2(\tau) \sim
\tau$.  The usual dispersion coefficient is thus defined in the
infinite time limit, $D = \lim_{\tau \to \infty} D(\tau)$.

For the OU process, the appropriate ensemble of particles is
a collection of independent realizations of the Wiener process.  
The diffusion coefficient then becomes 
\be
D_{ou}(\tau) = 2 \sigma_{ou}^2 T_L 
\left[1  - {{T_L \left(1 - e^{-\tau/T_L}\right)}\over {\tau}}\right],
\label{dispou}
\ee
which has the infinite time limit $D_{ou}= 2 \sigma_{ou}^2 T_L$.

For the point-vortex system, the appropriate ensemble in Equation
(\ref{dispdef}) is the collection of vortices or passives at a single
time in a single integration.  As for displacements, dispersion is
calculated using unfolded trajectories. The four resulting diffusion
coefficients are shown in Figure~\diffusion/.  There is no significant
difference between the vortices and passives, and thus the best
estimate is obtained by averaging the four diffusion coefficients.
The diffusion coefficient for the OU process, Equation (\ref{dispou}),
is shown for comparison.  Also shown is the expected error of the mean
for four ensembles of 100 OU particles, which is calculated with
standard statistical methods.  For short times, $t \lesssim 0.1
\sim T_L$, the OU process captures the dispersion reasonably well.
For intermediate times, $T_L \lesssim t \lesssim 2 \sim 20 T_L$, the
OU process overestimates the dispersion.  For long times, $t \gtrsim
20 T_L$, the diffusion coefficient appears to grow beyond that of the
OU prediction.  This anomalously large single-particle dispersion is
presumably due to the tails in the velocity pdfs and large
displacements of long-lived close dipoles. It is an interesting
question, which remains unanswered at the moment, whether the
single-particle dispersion becomes Brownian at very large times in
point-vortex systems.

\vfill\eject
\section{Conclusions}

In this paper we analyze the Lagrangian dynamics of vortices and
passives in high-dimensional, $N = O(100)$, point-vortex systems.  We
find that the behavior can be understood by partitioning the system
into two components: (1) mean-field behavior resulting from the
collection of distant vortices, and (2) high-velocity behavior
resulting from close pairs (and, occasionally, triplets or
more).

The partition in the statistical dynamics is due to the slow,
non-normal convergence of $p_N$.  In the limit $N \to \infty$ the
velocity pdf is Gaussian and only mean-field behavior occurs. The
mean-field behavior is well modeled as an Ornstein-Uhlenbeck (OU)
stochastic process.  This part of the dynamics has a short correlation
time, $T_L = 0.09$, which can be estimated by simple arguments based
on the turn-over times for average vortex separation.  However, due to
the $1/|\bu|^3$ behavior of the velocity pdf for a single vortex at
large $|\bu|$, the convergence to the Gaussian implied by the central
limit theorem is extremely slow with $N$, and the variance of the
Gaussian component scales as $\sqrt{N \ln N}$ rather than the more
common $\sqrt{N}$.

The slow convergence is important for physically relevant,
large but finite, values of $N$, where there is a significant,
non-Gaussian, high-velocity tail in the pdf.  This
high-velocity tail has a very long correlation time.  Close pairs
typically last for much longer than $T_L$; indeed, they can last so
long that even our longest integrations, of $O(300T_L)$, do not
yield reliable statistics for the distribution of pair lifetimes.
These long-lived pairs exhibit nonergodic behavior for at least the
time of our simulations, which is consistent with previous evidence
for long-time nonergodicity with only a few point
vortices.$^{\cite{Babiano94,WM91}}$

As a result of the high-velocity tails, the OU process does not
accurately capture the single-particle dispersion and the pdf of
particle displacements. The small displacements scale approximately as
in the OU model, but the large displacements show no such scaling due
to close dipoles. 
At very short times, of course, the displacements exhibit ballistic
scaling, with $D(\tau) \propto \tau$. After the first significant loss
of velocity correlation, at $t \sim O(T_L)$, the OU model captures the
average dispersion reasonably well.  At later times, $t \approx 1$ -
$5$, the OU model overestimates the single-particle dispersion, and at
very large times, $t > 20$, the OU model appears to underestimate the
dispersion. This suggests that the diffusion coefficient may continue
to increase resulting in long-time anomalous dispersion, rather than
saturate as in the OU model around a time of $O(10T_L)$.

It may be possible to construct a new stochastic process that includes
both the Gaussian slow vortices and the long-lived, close,
high-velocity pairs.  The slow vortices would be modeled as an OU
process.  Since the close same-sign pairs rotate in place, these
vortices should be modeled by a process that has negative velocity
correlations at very short times.  The close dipoles travel large
distances, and could be modeled by a L\'evy
walk.$^{\cite{Shlesinger}}$ In previous studies of point vortices,
Viecelli found anomalous relative dispersion at short times, which he
explained as L\'evy walks.$^{\cite{Viecelli90,Viecelli93}}$

In order to create a combined stochastic process one would need
information about the close-pair dynamics.  This information could be
condensed into two joint pdfs and two scalar probabilities: the joint
pdfs of close-pair velocities and lifetimes, separately for the
dipoles and same-sign pairs, and the transition probabilities for
switching from OU motion to close dipoles and same-sign pairs, i.e.,
the probabilities for close-pair formation.  We already know that the
pdf for close pair velocities, which is related to the joint pdf by
integrating over lifetime, scales as $u^{-3}$ for large $u$.  If we
are only concerned with displacements, then the same-sign pairs can
probably be absorbed into the OU component, since their center of
vorticity moves as a single vortex.  If we wish to model the
velocities, however, then the fast rotation of the same-sign pairs is
important.

With this information one could construct a stochastic algorithm in
which particles move with OU random walks, and, at particular instants,
have some probability of becoming members of a dipole or same-sign
pair.  Once a member of a dipole, the particle moves for some time at 
a constant velocity in a random direction, with the time and velocity
chosen from the joint pdf. Once a member of a same-sign pair, the
particle has a high oscillating velocity for some time, plus a slower
OU random-walk velocity, again with the time and velocity chosen from
the joint pdf.  Because of the long correlation time of the close pairs,
obtaining a good estimate of these joint pdfs would require significantly
longer integrations than we have done here; therefore, we will not
propose here an explicit form for the new stochastic model.

How relevant is the range of $N$ we study here for geophysical flows?
The number of coherent vortices is not well known in nature, since
they are often hard to detect by conventional measurement techniques;
for example, the vorticity or potential vorticity fields are usually
poorly sampled.  However, we can estimate an upper bound on the number
of coherent vortices in the large-scale geostrophic regimes of the
atmosphere and ocean as the ratio of total area to a typical vortex
area, neglecting any stacking in the vertical.  (We do not consider
the ageostrophic, smaller-scale regimes, for which 2D vortex dynamics
are probably less relevant.) A typical geostrophic vortex size is
given by the Rossby radius of deformation, which, at midlatitudes, is
approximately $30$ km in the ocean, and $1000$ km in the
atmosphere.$^{\cite{Gill}}$ The total areas of the ocean and atmosphere
are about $3.5\times10^8$ km$^2$ and $5\times10^8$ km$^2$,
respectively.  Thus, if the area were filled with closely packed
vortices, which is certainly not the case, the ocean would have
$O(10^5)$ vortices and the atmosphere would have only $O(10^2)$
vortices.  From Figure \cenlimpv/ we see that with this upper bound
even the ocean would not be in the infinite $N$ regime and the tails
of the pdf would be important.  Thus, the behavior described in this
paper is representative of what we expect from the number of coherent
vortices in the atmosphere and ocean.

The point-vortex system is only an approximation to the dynamics of
coherent vortices in geophysical flows. True coherent vortices have a
finite core size and a finite lifetime, even though the former is
often small and the latter large compared to many geophysical
phenomena. The existence of a finite core does not significantly
affect the dynamics of passive tracers, as shown by simulations of
two-dimensional turbulence.$^{\cite{EPB93,Babiano94,Provenzale97}}$
Since extended coherent vortices effectively retain the matter in and
near their cores and can travel long distances over their long
lifetimes, they can have considerable influence over large-scale
material transports, similarly to what happens for point vortices.  On
the other hand, the dynamics of the vortices themselves may be
affected by the presence of a finite core. In particular, same-sign
vortices can merge, a process which is not included in point-vortex
dynamics.  Although a full answer to the impact of finite cores can
only come from simulations of extended vortices, we note that many of
the specific Lagrangian properties discussed here are mainly
determined by the behavior of close dipoles. These still exist, with
similar properties, in the case of extended vortices, suggesting that
the results found here for point vortices which are due to dipoles may
be more general. Future studies will address the properties of
Lagrangian transport in systems of coherent vortices with extended
cores and in punctuated point vortex dynamics where instantaneous
merger is allowed.

In this paper we have demonstrated several differences of vortex
systems from random walking.  On the other hand, aspects of the overall
dispersion behavior seen in Figure~\diffusion/ are grossly captured by
simple OU diffusion on times very large compared to $T_L$.  Thus, we
see this study as giving some further degree of support for the common
practice of parameterizing large-scale transports as eddy diffusion.
In this context, the eddy diffusion coefficient is a function of the
coherent vortex population obtained from Equation (\ref{dispou}) with
$\sigma_{ou}$ replaced by $\sigma_N$, $D = N T_L \ln N/\pi$.  Thus,
one may be able to obtain a time-varying eddy diffusion by observing
how the vortex population changes over time.  However, it may
eventually turn out that the quantitatively significant differences
between vortex dispersion and stochastic diffusion imply a significant
qualitative difference as well.  For example, if the diffusion 
coefficient continues to grow with time, then for times much longer
than these simulations, eddy diffusion will fail.

\section{Acknowledgments}

The numerical computations of this work have been carried on at the
computing center of the Istituto di Cosmogeofisica, Torino, Italy, and
at the Laboratory for Computational Dynamics of JILA, University of
Colorado, Boulder. JW is partially supported by NSF ECS-9217394 and
DOC NA56GP0230. AP is grateful to JILA for support and hospitality
during the final part of this work.  Support from the EC contract
EV5V-CT94-0503 is gratefully acknowledged.

\appendix

\section{Appendix:  The Central Limit Theorem for Point Vortices}

In this appendix we show that in the limit $N\to\infty$ the velocity
probability density function (pdf) for the velocity due to $N$
randomly placed vortices, $p_N$, becomes Gaussian with variance
$\sigma_{N} = \sqrt{N \ln N/2 \pi}$.  The calculation is based on
theorems regarding random variables.$^{\cite{Ibrag}}$ The fact that
$p_N$ becomes Gaussian was mentioned previously by Min, et
al;$^{\cite{Min96}}$ the value of the variance is, to our knowledge,
new.

The relevant theorems discuss the behavior of $z_N$, the sum of $N$
independent identically distributed random variables $x_i$ with pdf
$F(x)$, scaled by a
factor $\sigma_N$: 
\be
z_N = {{x_1 + \ldots + x_N}\over{\sigma_N}}.
\ee
If in the limit $N \to \infty$ the pdf for $z_N$ converges weakly to
some distribution $G$, then $F$ is said to be in the domain of
attraction of $G$.  The family of distributions which have non-empty
domains of attraction are called L\'evy distributions and are
identified by an index $\alpha,$ $0 < \alpha \le 2$; $\alpha = 2$
corresponds to the normal distribution, and $\alpha = 1$ is the Cauchy
distribution.$^{\cite{Ibrag,Shlesinger}}$  Here we shall only be
concerned with $\alpha = 2$.

Ibragimov and Linnik$^{\cite{Ibrag}}$ prove that $F$ belongs to the domain of
attraction of the normal distribution if and only if the truncated
second moment  
\be
I(X) = \int_{-X}^X x^2 F(x) \,dx
\label{Izdef}
\ee
is slowly varying in the sense that 
\be
\lim_{X\to\infty} {{I(t X)}\over{I(X)}} = 1
\label{slow}
\ee
for all $t$.  They further show that the scaling of the sum,
$\sigma_N$, is given by the requirement that 
\be
\lim _{N\to\infty} {{N I(\eps \sigma_N)}\over {\sigma_N^2}}  = 1,
\label{Bnreq}
\ee 
for some $\epsilon > 0$. As written, Equation (\ref{Bnreq}) assumes that
$F$ has zero mean which is the case of interest here; if $F$ has
nonzero mean then another term is needed in the equation.  

Here, $p_1(u)$ takes the role of $F(x)$, where $p_1(u)$ is the pdf for
the $x$-component of the velocity, $u$, produced at the origin from a
point vortex with unit circulation placed at a random position in the
periodic domain $[-\pi,\pi]^2$.   Since the velocity is a vector $\bu
= (u,v)$ described by a joint distribution $p(u,v)$, $p_1$ is obtained
from   
\be
p_1(u) = \int_{-\infty}^{\infty} p(u,v) \,dv.
\ee
Due to symmetry the pdfs for $u$ and $v$ are identical and we need
only consider one of them.  It is sufficient to only consider vortices
with positive circulation since changing the sign of the circulation
is equivalent to changing the sign of $\bx$. Thus, allowing random
positive and negative unit circulations would not affect $p_1$ or
$p_N$. 

We are interested in the velocity component $u$ due to $N$ vortices,
with pdf $p_N(u)$, where $u$ is the unscaled sum of the velocity
components of the individual vortices.  Thus, if the sum scaled by
$\sigma_N$ has a normal distribution, then $p_N$ is Gaussian with
variance $\sigma_N$. 

While the joint distribution $p(u,v)$ is rather difficult to
determine in detail, changing variables allows us to easily integrate
over the distribution.  The random velocity $\bu$ is due to
a vortex at a random position $\bx$, so $p(u,v)\,du \,dv =
p(x,y)\,dx\, dy$.  Further, since the vortex is
placed randomly from a uniform  distribution, $p(x,y)=1/4\pi^2$.  A
final change of variables to polar coordinates $(r,\theta)$ gives 
\ba
I(X) &=& \int_{-X}^X du \, \int_{-\infty}^\infty dv\, u^2 p(u,v),  
\nonumber\\
&=& {{1}\over{4 \pi^2}} \int d\theta \, \int dr \,r\, u^2(r,\theta).
\ea

To determine the bounds of the integral we need the curve
$r(\theta,X)$ such that $u(r,\theta)
= X$, with $v$ arbitrary.  Since we are interested in the behavior of
$I(X)$ in the limit of large $X$ we can use the asymptotic form of
Equation (\ref{vel}), $u(r,\theta) = 2 \sin\theta/r$.  Since the
original integral is over $|u| < X$, in polar coordinates the
integration is over $r > 2\sin\theta/X$; this corresponds to
integrating over all points outside two circles with radius $1/X$
centered at $\pm 1/X$.  Along the $x$-axis, $\theta = 0$ or $\pi$, $r$
is allowed to go to zero; thus, $v = 2\cos\theta/r$ can reach both $\pm
\infty$ and $r > 2\sin\theta/X$ does not restrict $v$. Defining
$R(\theta)$ to be the distance from the origin to the edge of the
periodic domain results in
\be
I(X) = {{1}\over{4 \pi^2}} \int_0^{2 \pi} d\theta \, 
\int_{2\sin\theta/X}^{R(\theta)} dr \,r \,u^2(r,\theta).
\ee

Despite the fact that the integral covers small $u$, the limiting
behavior of $I$ only depends on the asymptotic form of $u(r,\theta)$.
Writing the velocity as
\be
u(r,\theta) = {{2 \sin\theta}\over{r}} + g(r,\theta),
\ee
where, from Equation (\ref{vel}), $g \sim O(r)$ for small $r$, gives  
\be
I(X) = {{1}\over{\pi}} \ln X + O(1).
\ee
It is now straightforward to verify that $I(X)$ is slowly varying in
the sense of (\ref{slow}), and thus $p_1$ is in the domain of
attraction of the normal distribution.  Further, the requirement
(\ref{Bnreq}) is satisfied by 
\be
\sigma_N = \sqrt{{N \ln N}\over{2 \pi}}.
\ee
Thus, in the limit $N\to\infty$, $p_N$ is Gaussian with variance
$\sigma_N$.  Because this result only depends on the explicit form of
$u$ for small separations, it is independent of boundary conditions.

\vfill\eject

\centerline{\bf Figure Captions}

\bigskip

Figure~\cenlimpv/. Probability density function $p_N(u/\sigma_N)$ for
the instantaneous velocity $u$ produced by $N$ randomly placed
vortices, scaled by $\sigma_N = \sqrt{N \ln N / 2 \pi}$.  Solid lines
are $p_N$ for $N=10^2$, $10^4$, and $10^6$.  Dashed line is a Gaussian
pdf with unit variance and amplitude 0.9.  The inset shows same curves on
log-log axes, along with the dotted line $\sim 1/u^3$.

Figure~\denstate/. Density of states as a function of energy for $10^4$
random configurations of $100$ vortices.  Dashed lines indicate two
arbitrary initial condition used for long integrations.

Figure~\vrscatt/. Scatter-plot of instantaneous vortex speed 
versus nearest neighbor distance for all vortices in $300$ randomly
chosen configurations. The dashed lines indicate the speed induced
by a single vortex with the smallest and largest circulations in the
population. 

Figure~\longpdf/. The velocity pdf averaged over time and particles,
$\bar p(u)$, for a) the vortices, and b) the passives.  In each panel,
the two solid lines are $\bar p(u)$ for the two long integrations
discussed in the text, the dotted line is a Gaussian with variance
$\sigma_{100}$ and amplitude 0.9 and the dashed line is $p_{100}(u)$,
as in Figure~\cenlimpv/, but with the same sample size as $\bar p$.

Figure~\autocorr/. Velocity autocorrelation function $R(\tau)$, Equation
(\ref{autodef}), for the two long integrations.  Solid lines are the
vortices, dashed lines are the passives. The dotted line is the
exponential which best fits the average of the vortices and passives,
and has a decay time of $T_L = 0.09$.

Figure~\traj/. Individual particle trajectories from the long
integrations: a) a fast vortex, where the period of fast motion,
$|\bu| > 70$, is shown in bold, b) a slow vortex, c) a slow passive,
and d) a stochastic particle governed by the Ornstein-Uhlenbeck
process, Equation (\ref{ou}).  The trajectory of the fast vortex
appears not to be smooth because of the finite plotting interval,
$\Delta t_s$.

Figure~\vtfast/. Time series for the speed of the fast vortex shown in
Figure~\traj/(a). 

Figure~\nnfast/. Nearest neighbor distance for the fast vortex shown
in Figure~\traj/(a) (lower curve).  The identity of the nearest
neighbor is shown in the upper curve: a line at 1.3 indicates a
same-sign nearest neighbor, a line at 1.1 indicates an opposite-sign
nearest neighbor, and the line jumps to 1.2 every time the
identity of the nearest neighbor changes.

Figure~\distratio/. Ratio of the distances from the fast vortex shown
in Figure~\traj/(a) to its second and first nearest neighbors.

Figure~\vordisp/. Probability density function of displacements $\D x
(\tau)$ scaled by the OU expected displacement $\sigma_{\D x}$, 
Equation (\ref{Dxou}), for $\tau = 0.1, 0.2, \ldots, 0.9$, from the
$300$ short integrations.  The dashed line is the OU Gaussian.  The inset
shows the tails for $\tau=0.1$ (upper solid curve) and $0.9$ (lower
solid curve). 

Figure~\drboth/. Scatter-plot of displacement $\D x (\tau)$, $\tau =
0.05$, versus average nearest neighbor distance over the time period
for a) opposite-sign pairs, and b) same-sign pairs in the $300$ short
integrations.  The solid line in a) is the displacement of a single
dipole on the infinite plane with $|\G|=1$.

Figure~\diffusion/. Diffusion coefficient $D(\tau)$ from the two long
integrations for the vortices (thin solid curves), passives (thin
dashed curves), and the average of the four (thick solid curve).  Also
shown is the OU diffusion coefficient (thick solid smooth curve) and
the envelope of expected error for four ensembles of $100$ OU
particles (thick dashed curves).  The inset shows the same curves for
short times.

\vfill\eject

\begin{figure}
\epsfbox{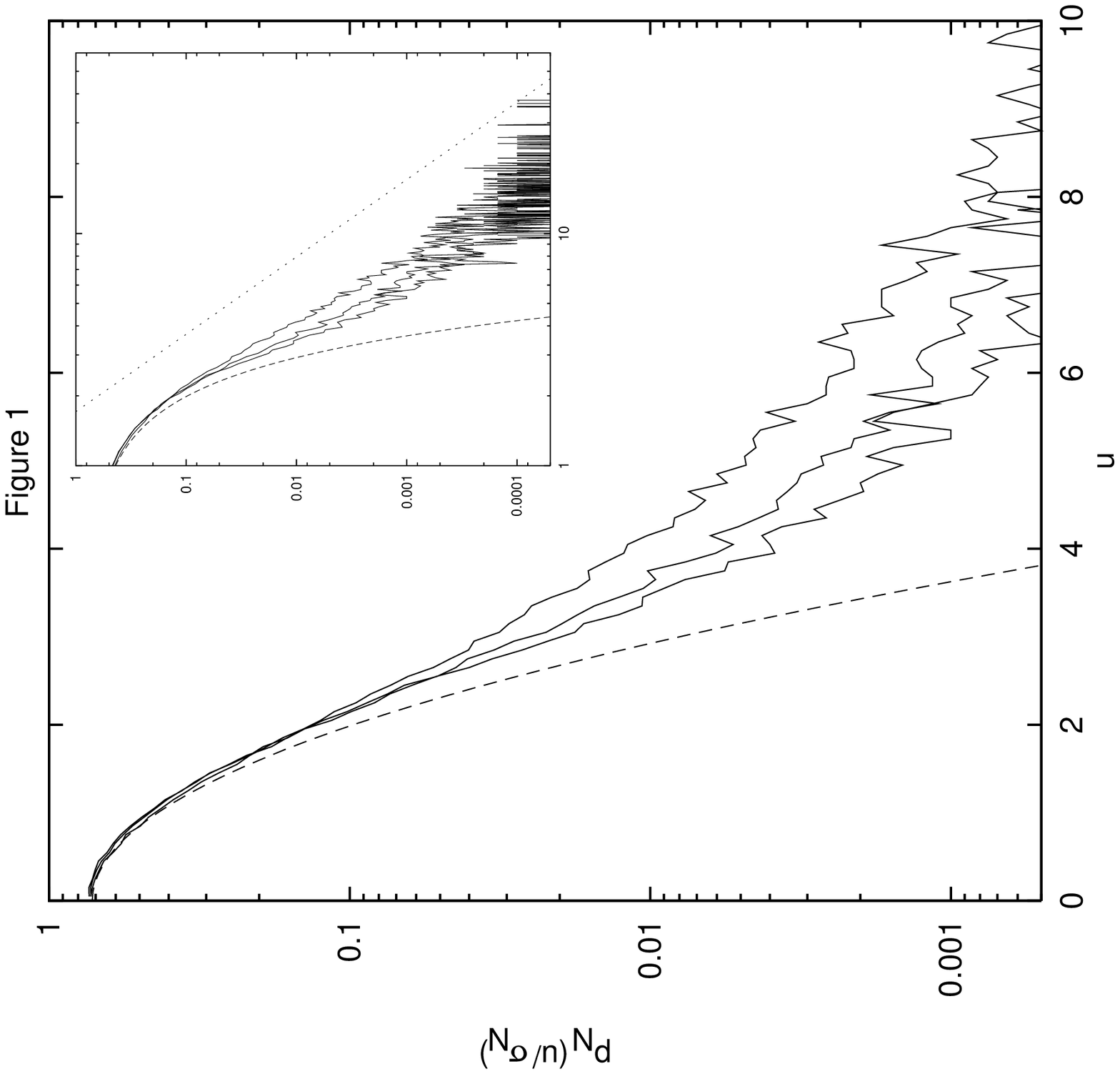}
\end{figure}

\begin{figure}
\epsfbox{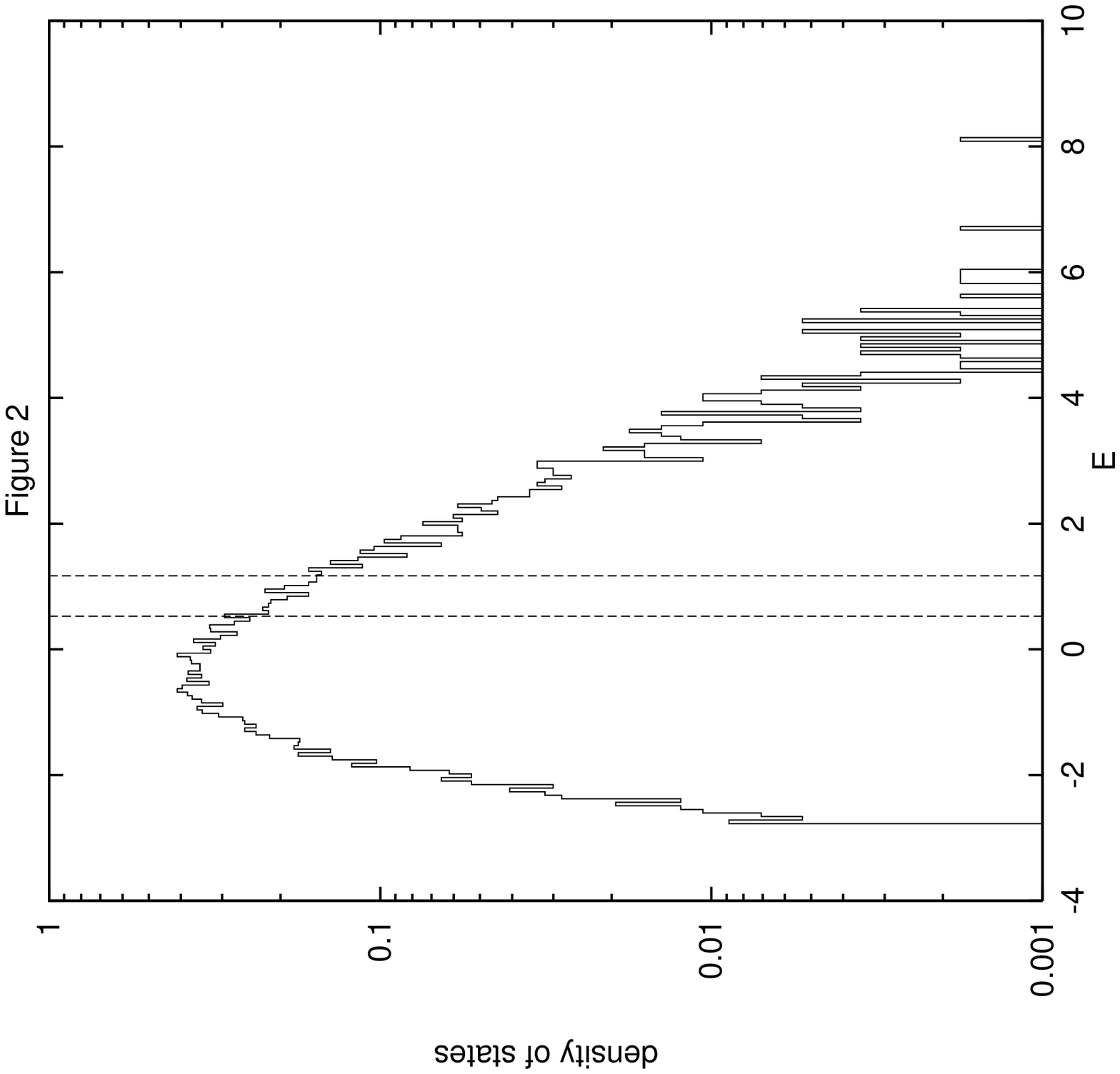}
\end{figure}

\begin{figure}
\epsfbox{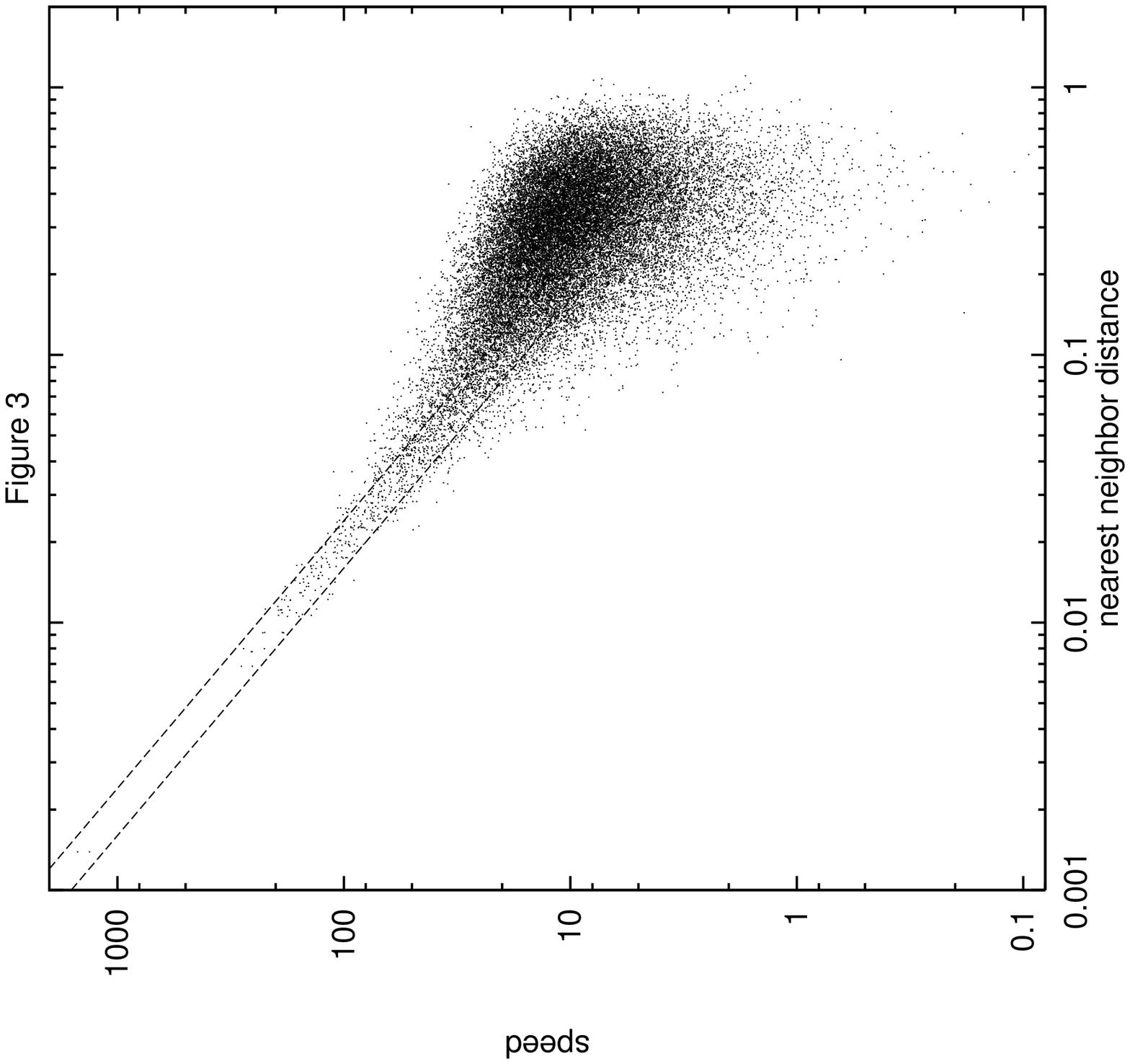}
\end{figure}

\begin{figure}
\epsfbox{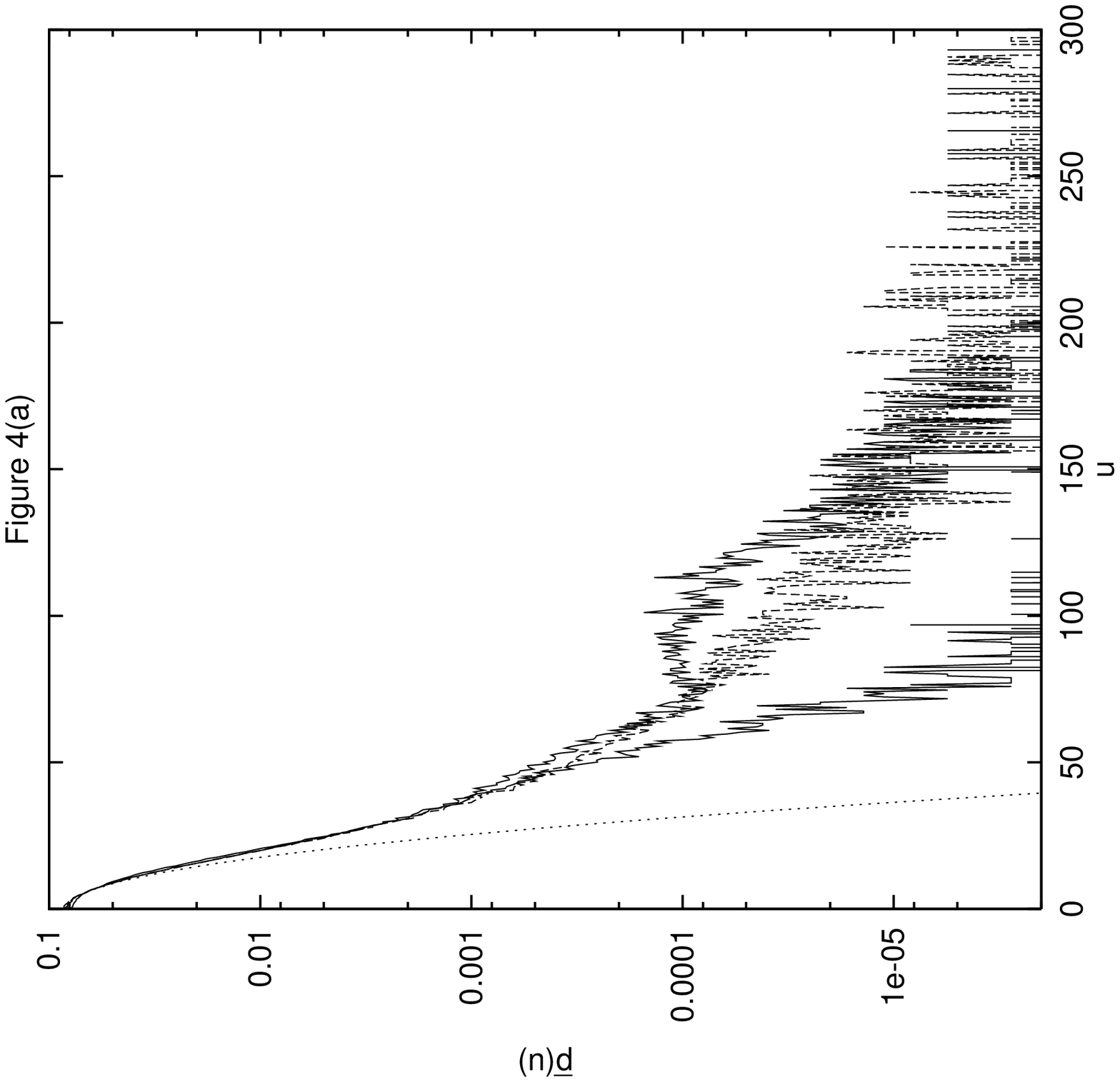}
\end{figure}

\begin{figure}
\epsfbox{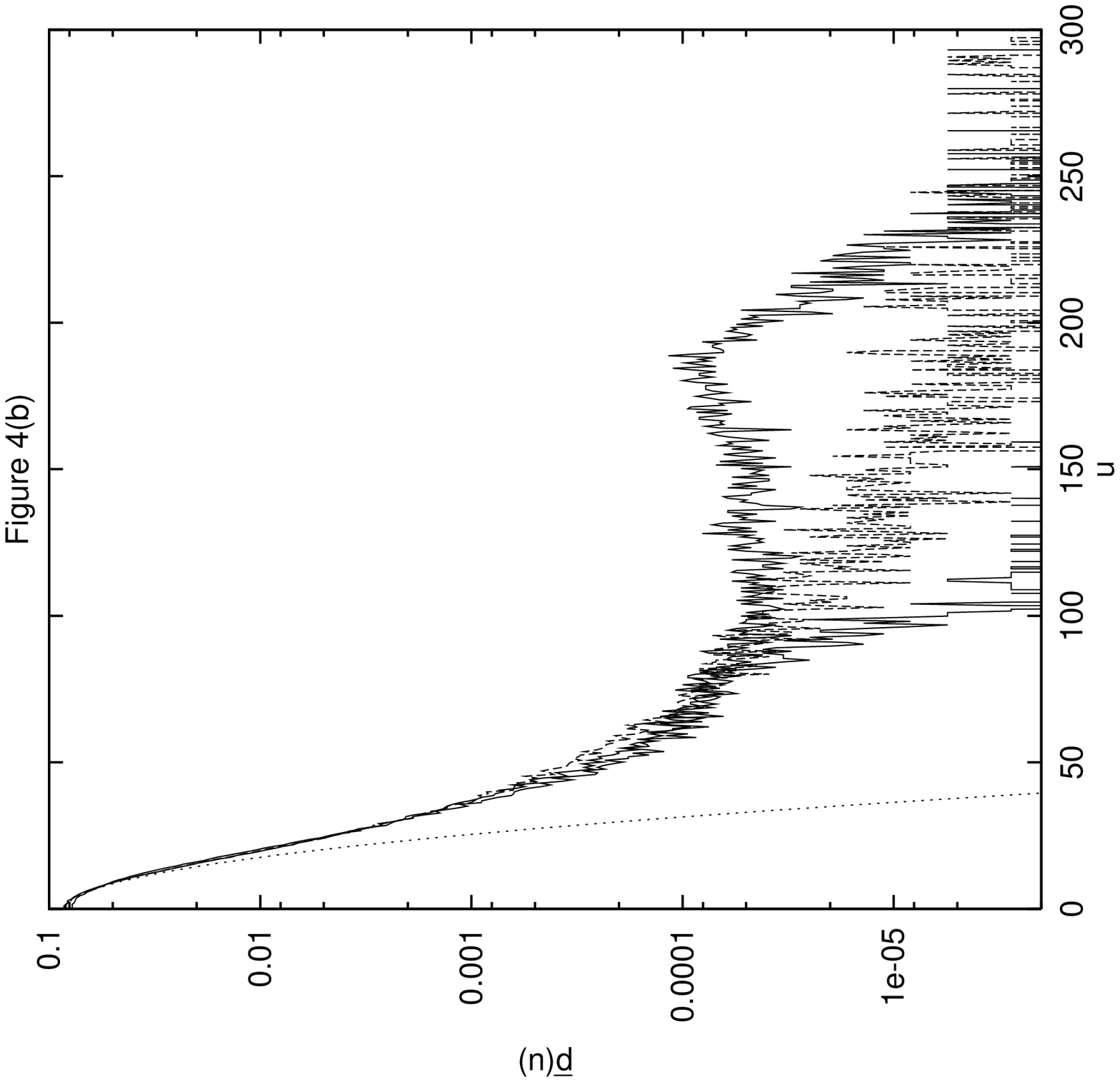}
\end{figure}

\begin{figure}
\epsfbox{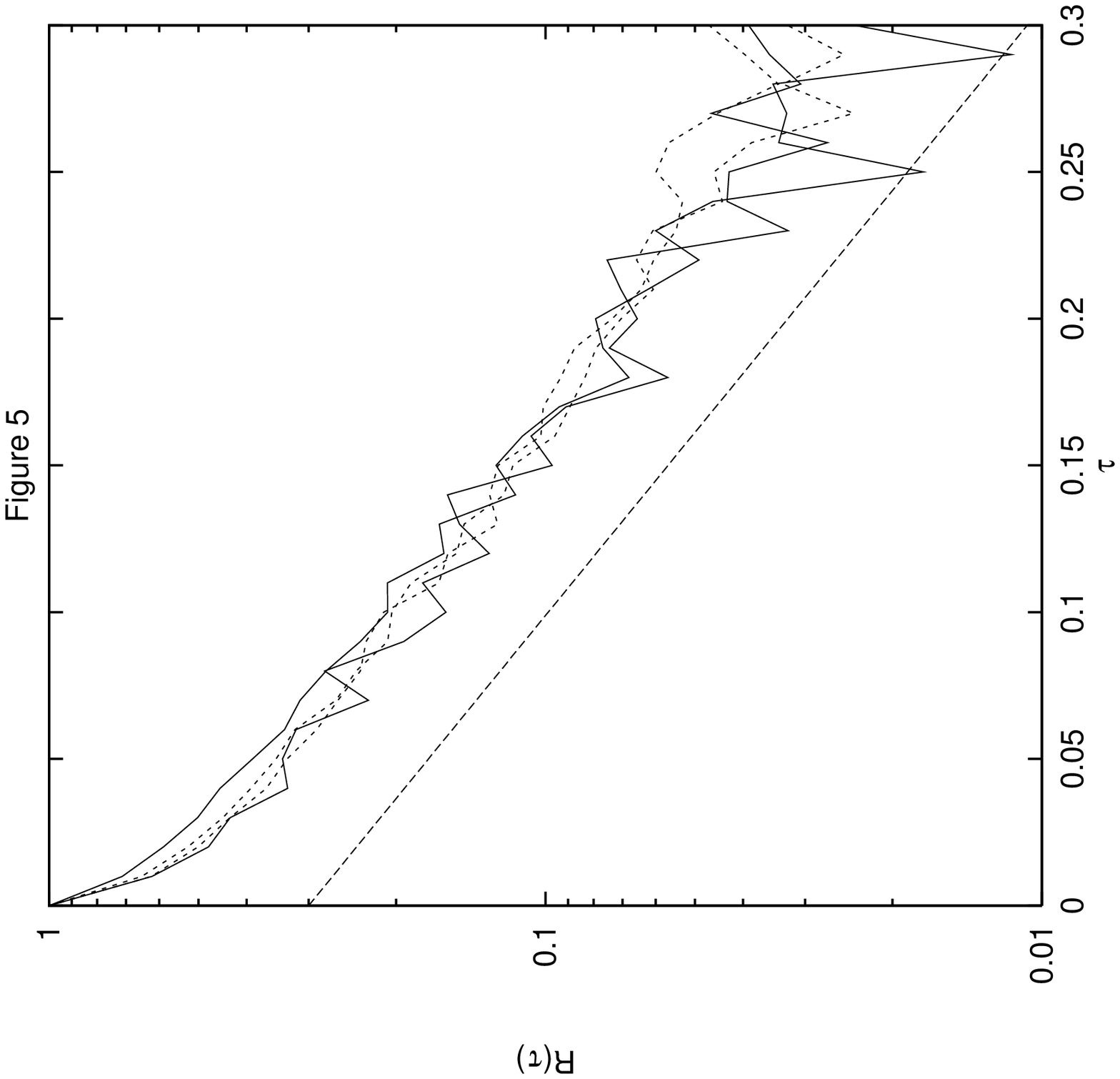}
\end{figure}

\begin{figure}
\epsfbox{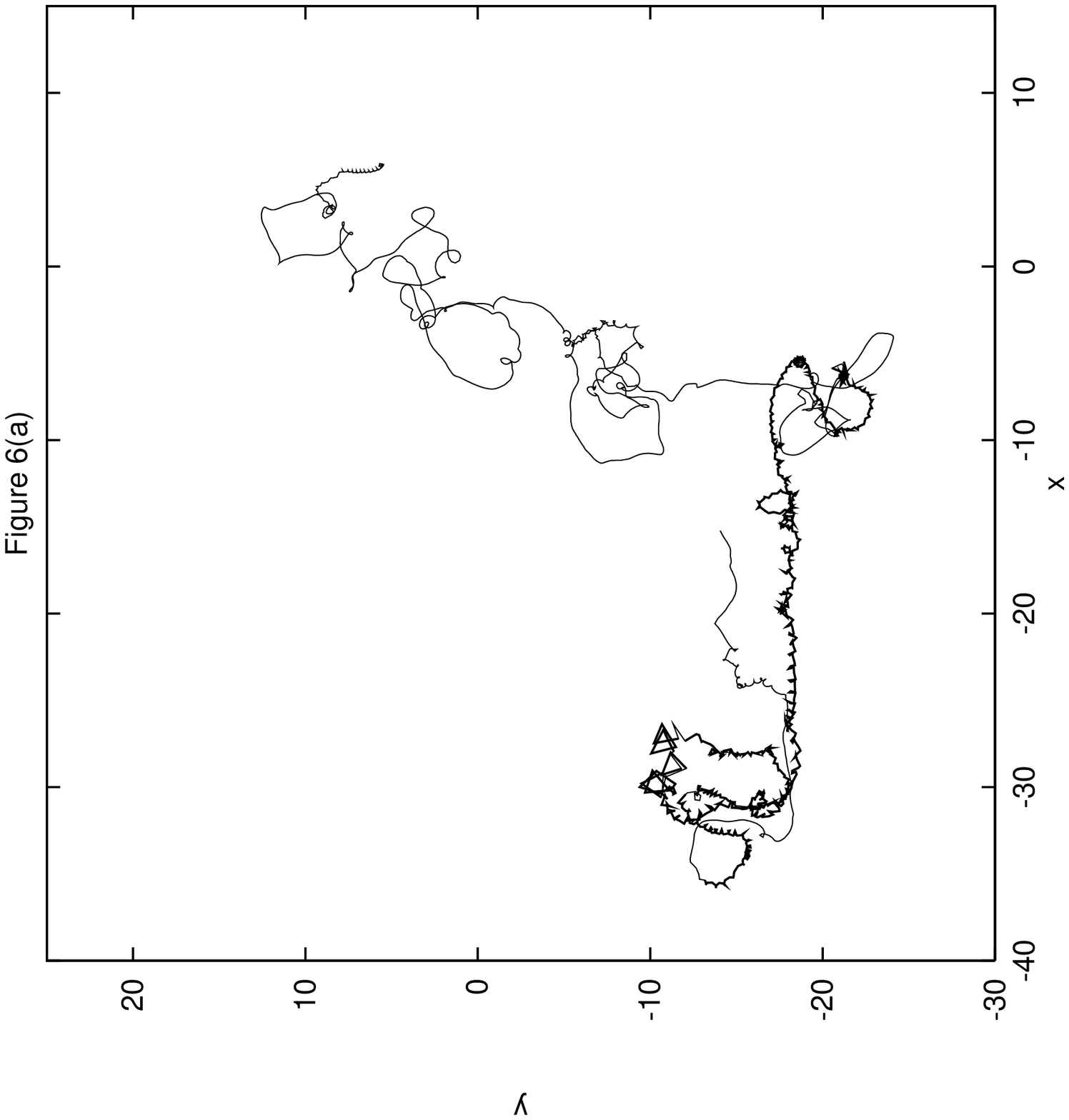}
\end{figure}

\begin{figure}
\epsfbox{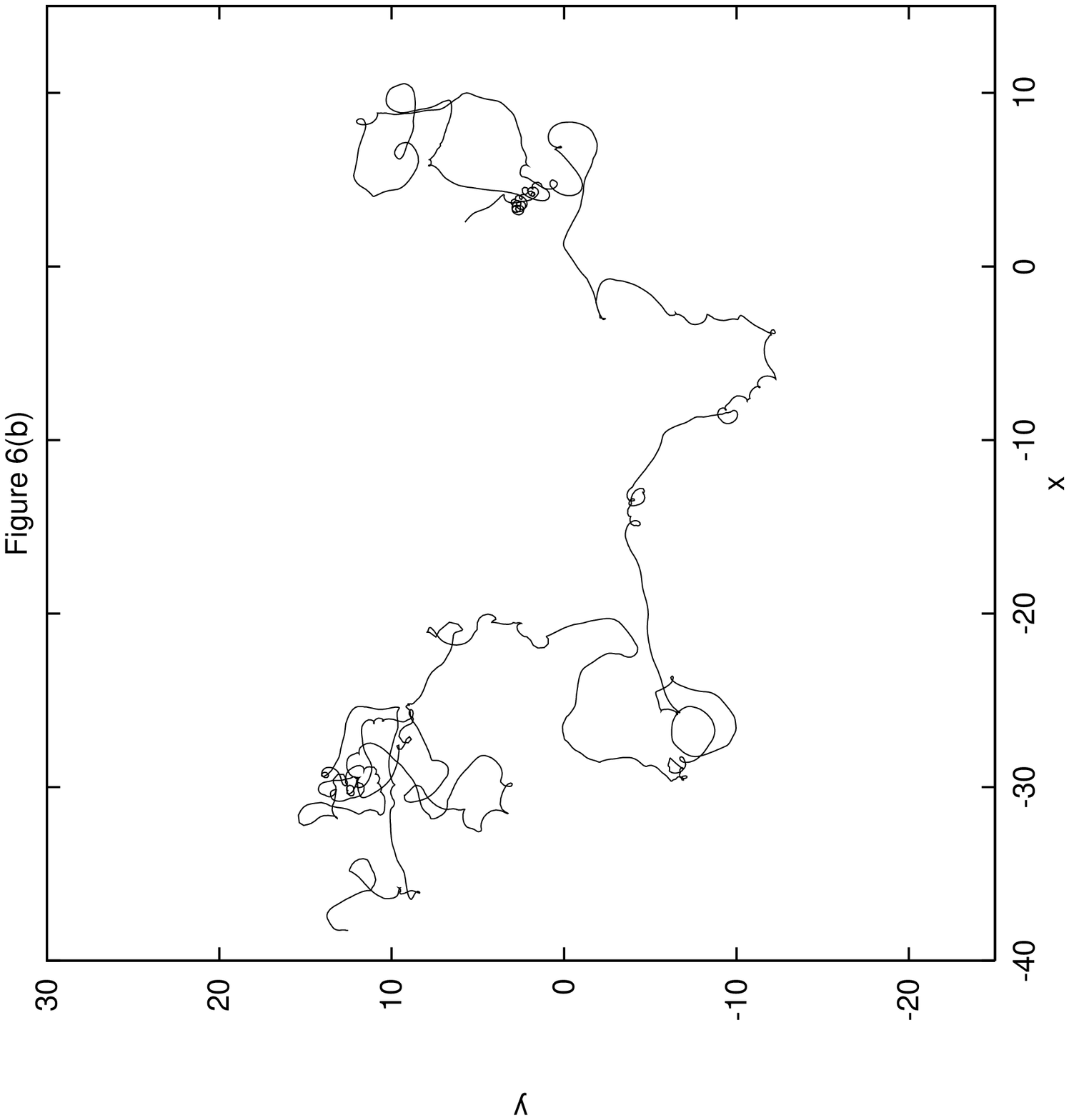}
\end{figure}

\begin{figure}
\epsfbox{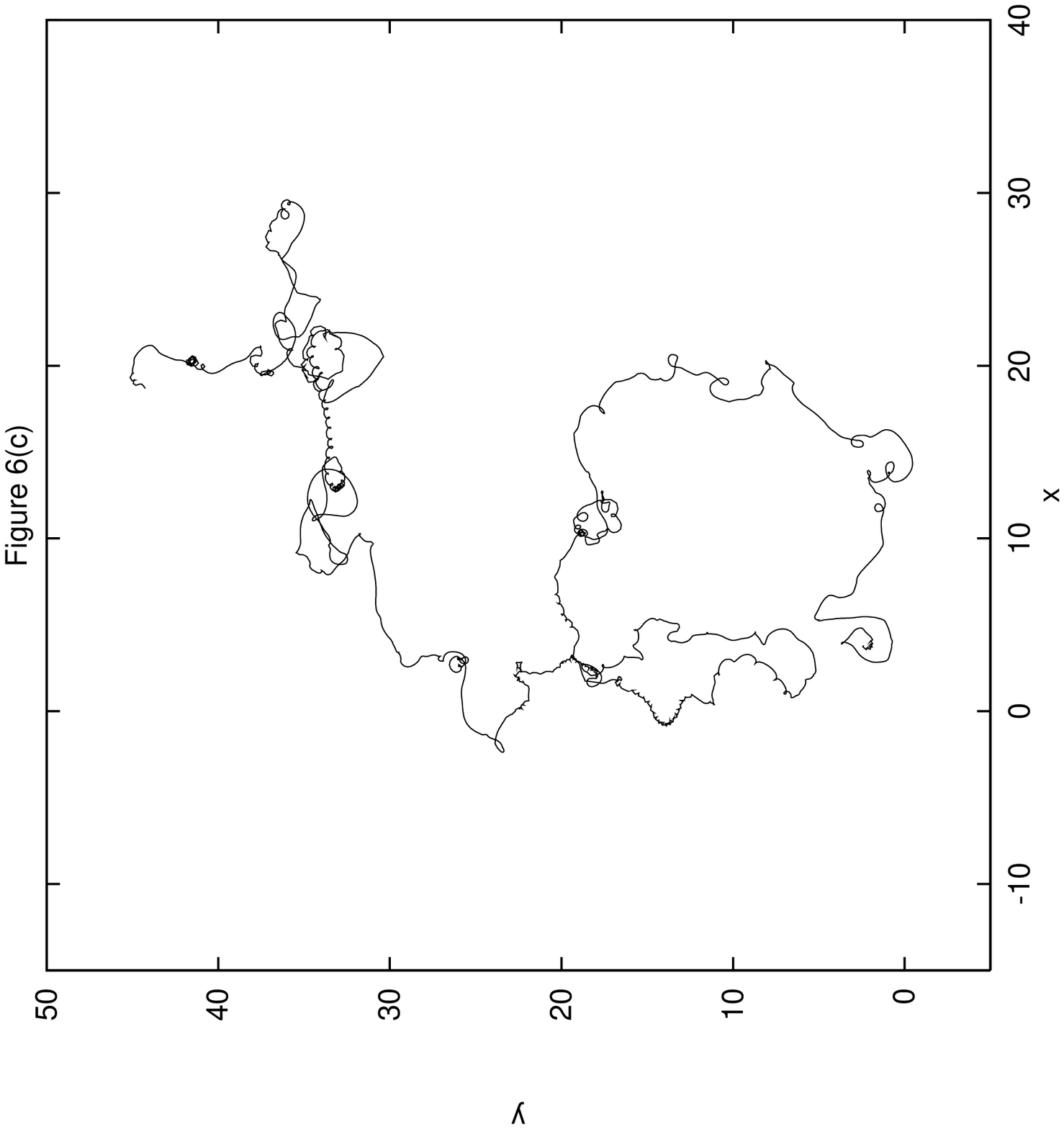}
\end{figure}

\begin{figure}
\epsfbox{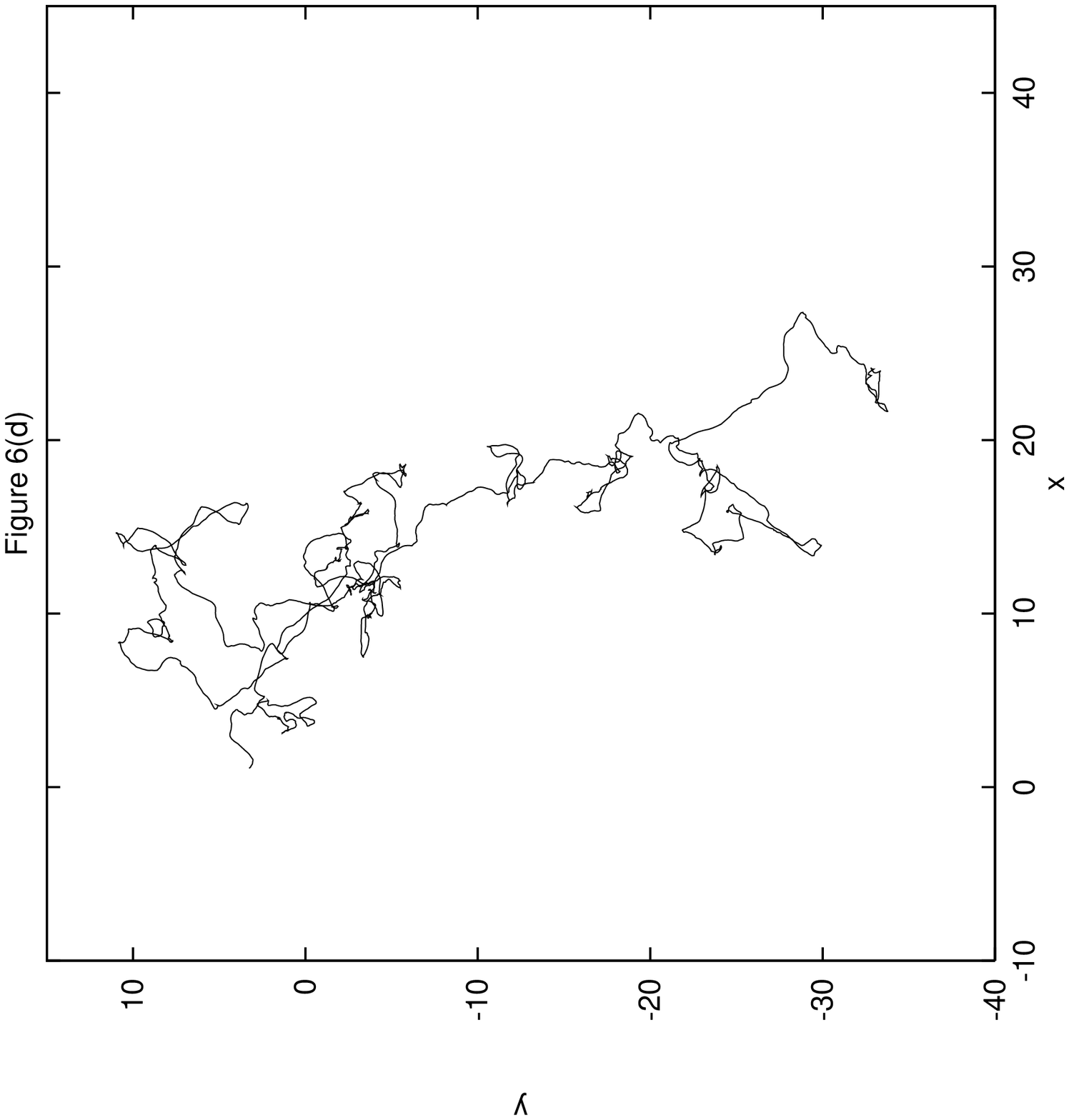}
\end{figure}

\begin{figure}
\epsfbox{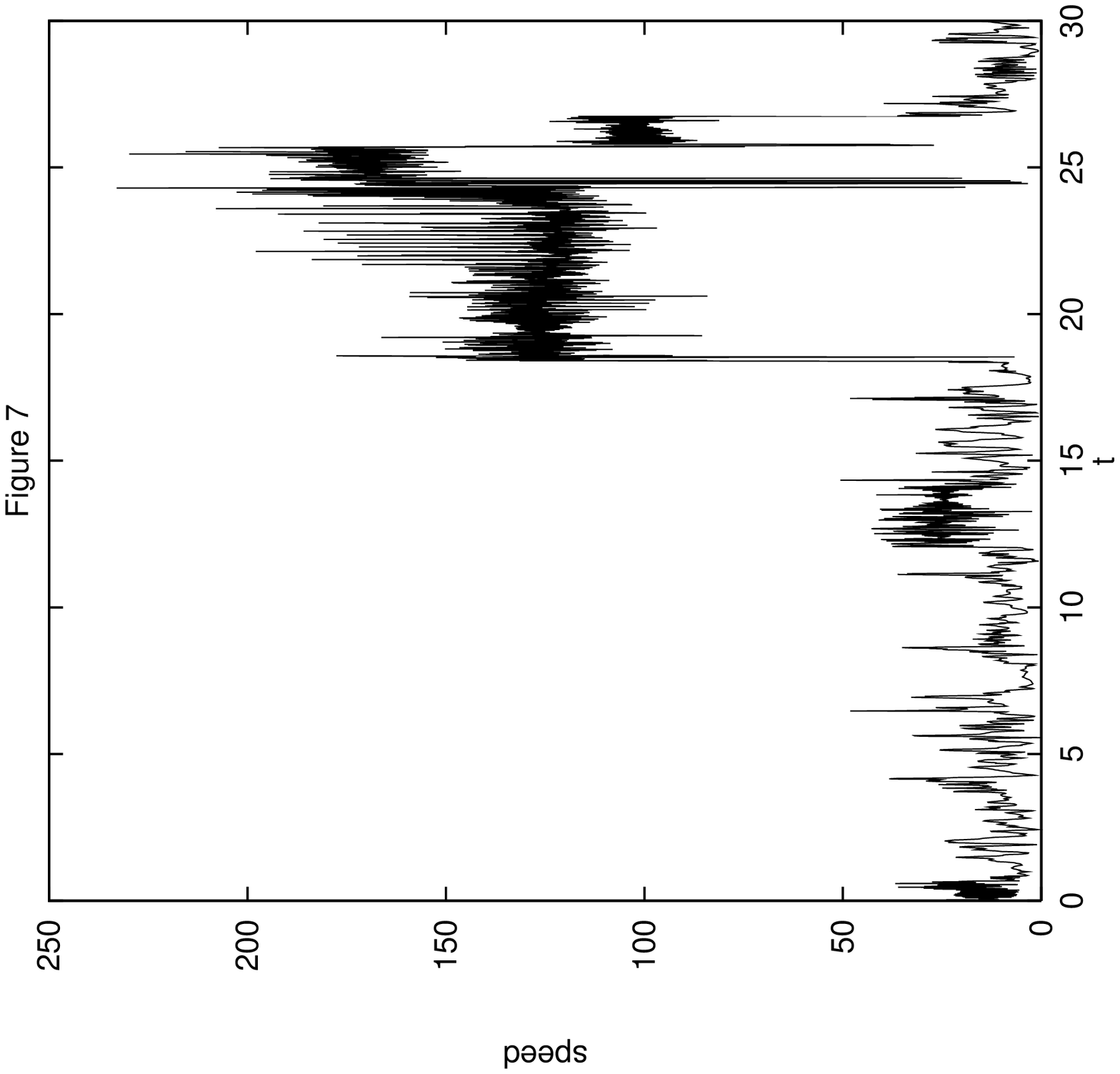}
\end{figure}

\begin{figure}
\epsfbox{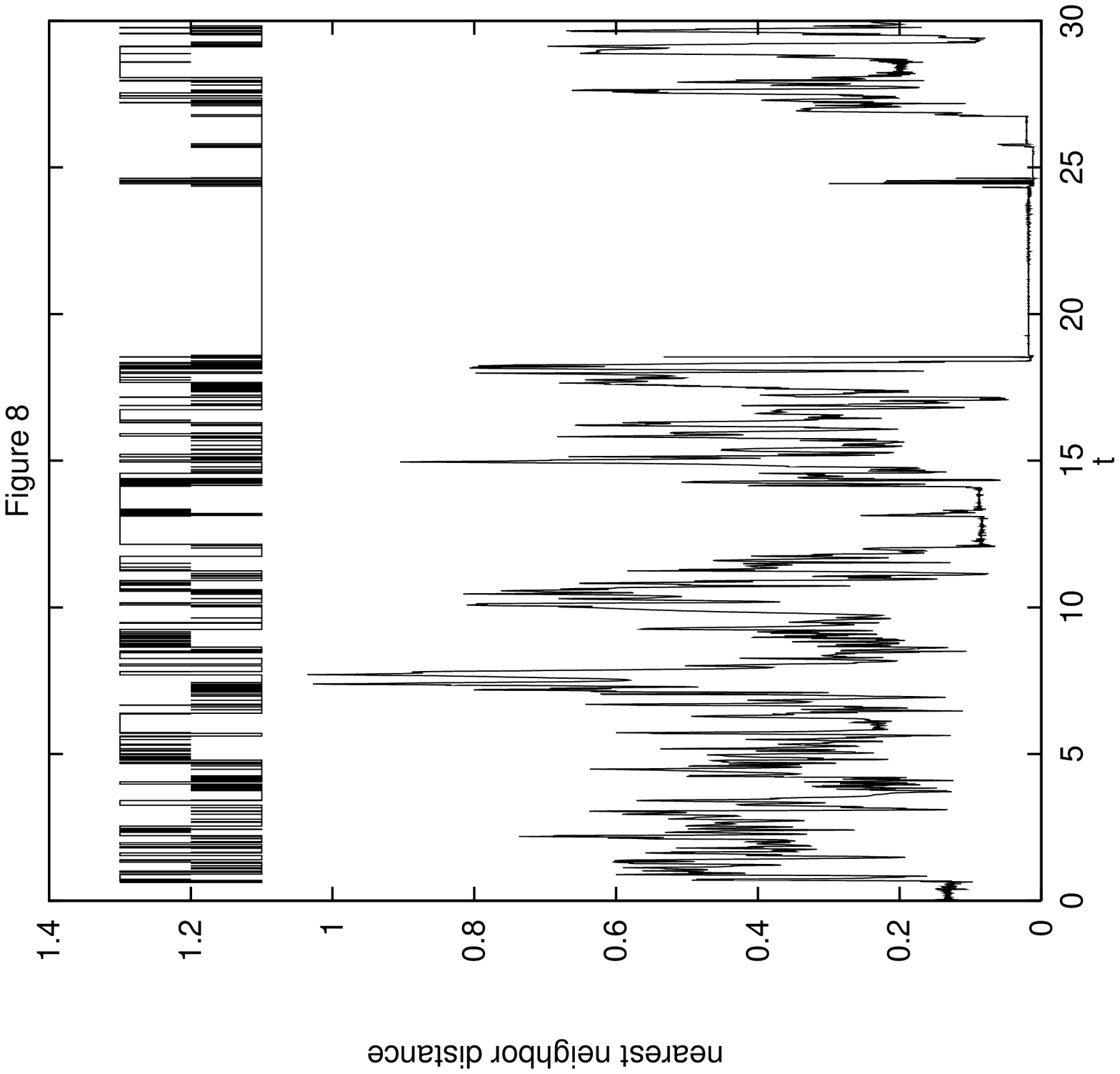}
\end{figure}

\begin{figure}
\epsfbox{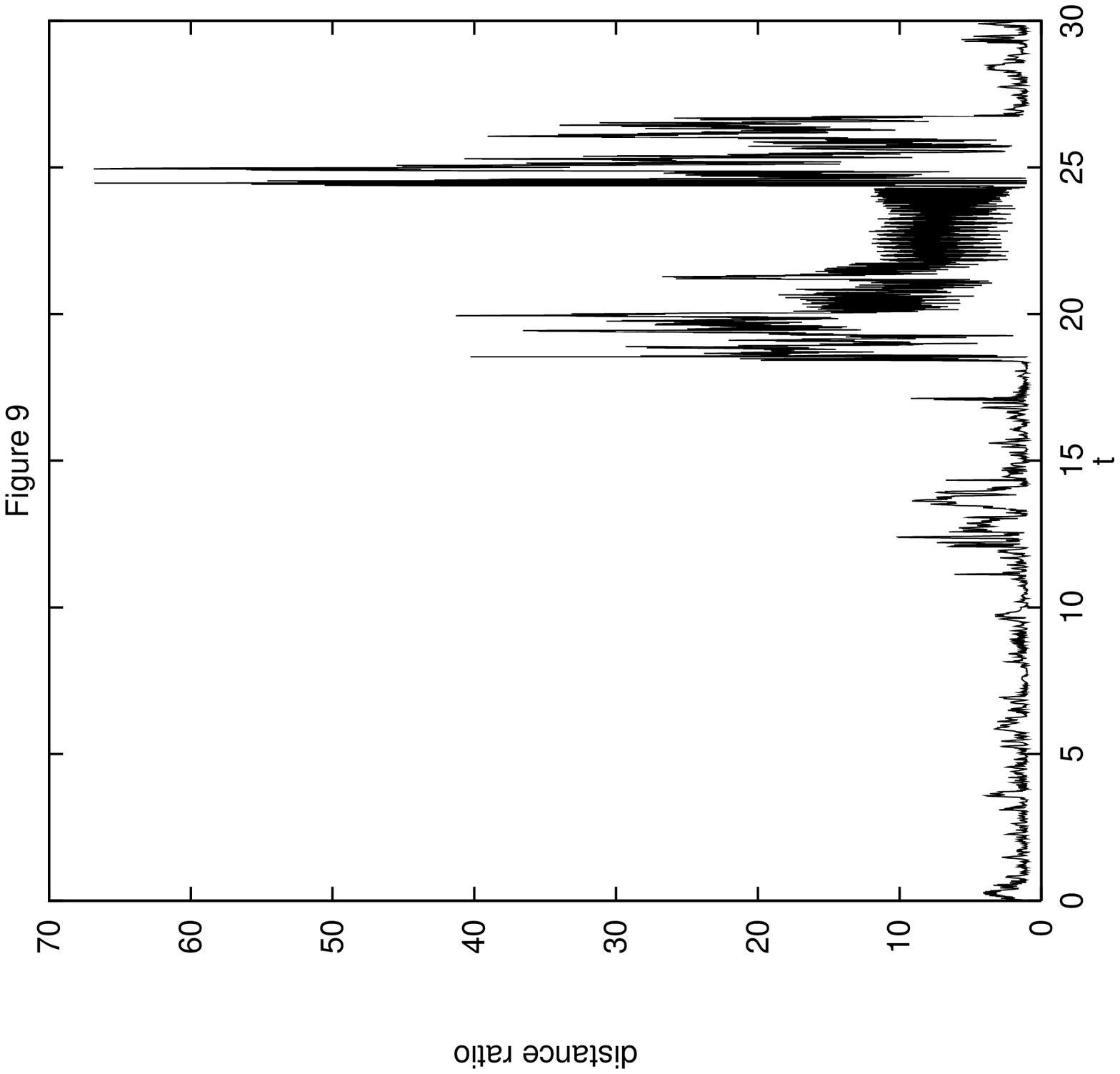}
\end{figure}

\begin{figure}
\epsfbox{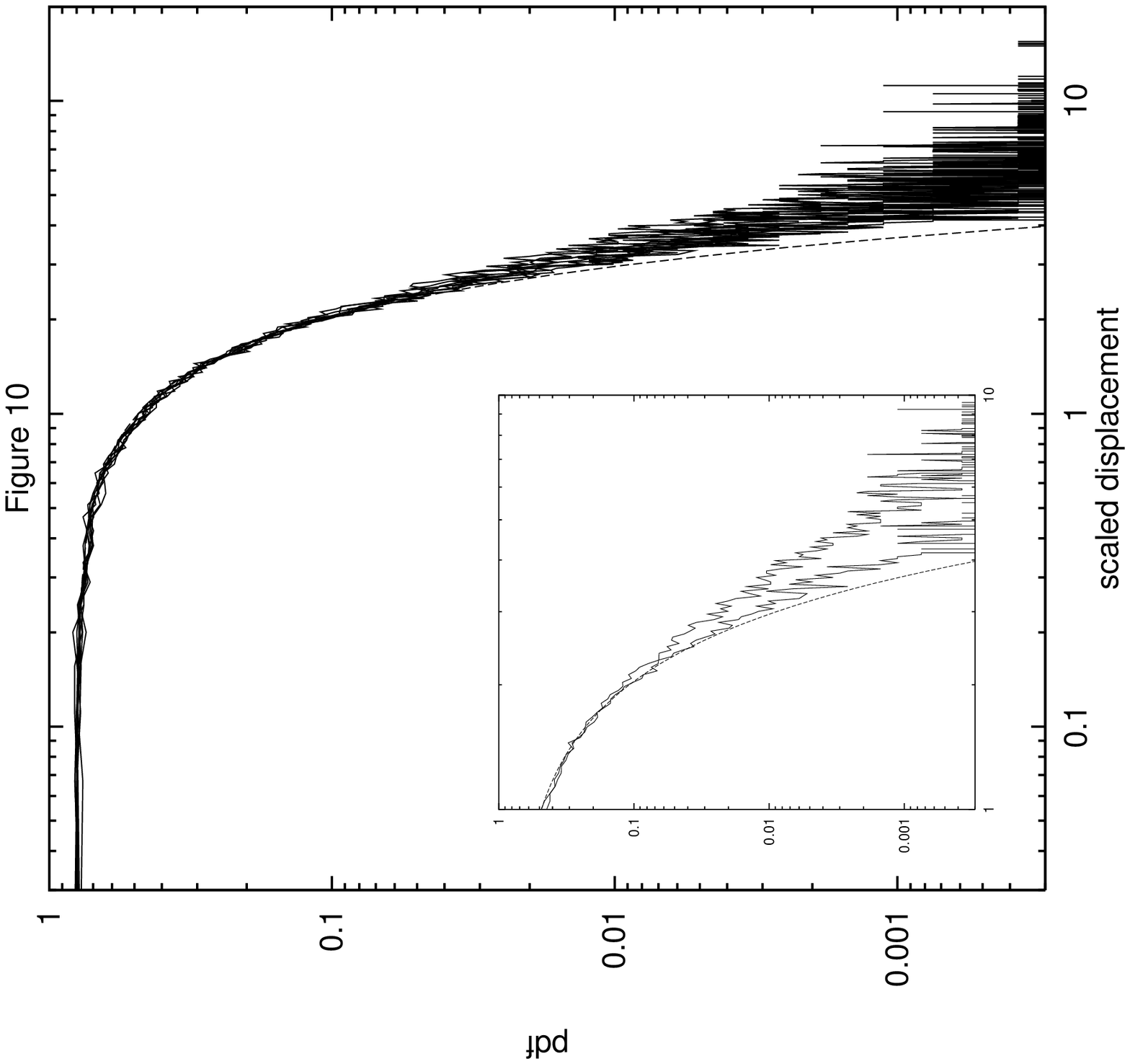}
\end{figure}

\begin{figure}
\epsfbox{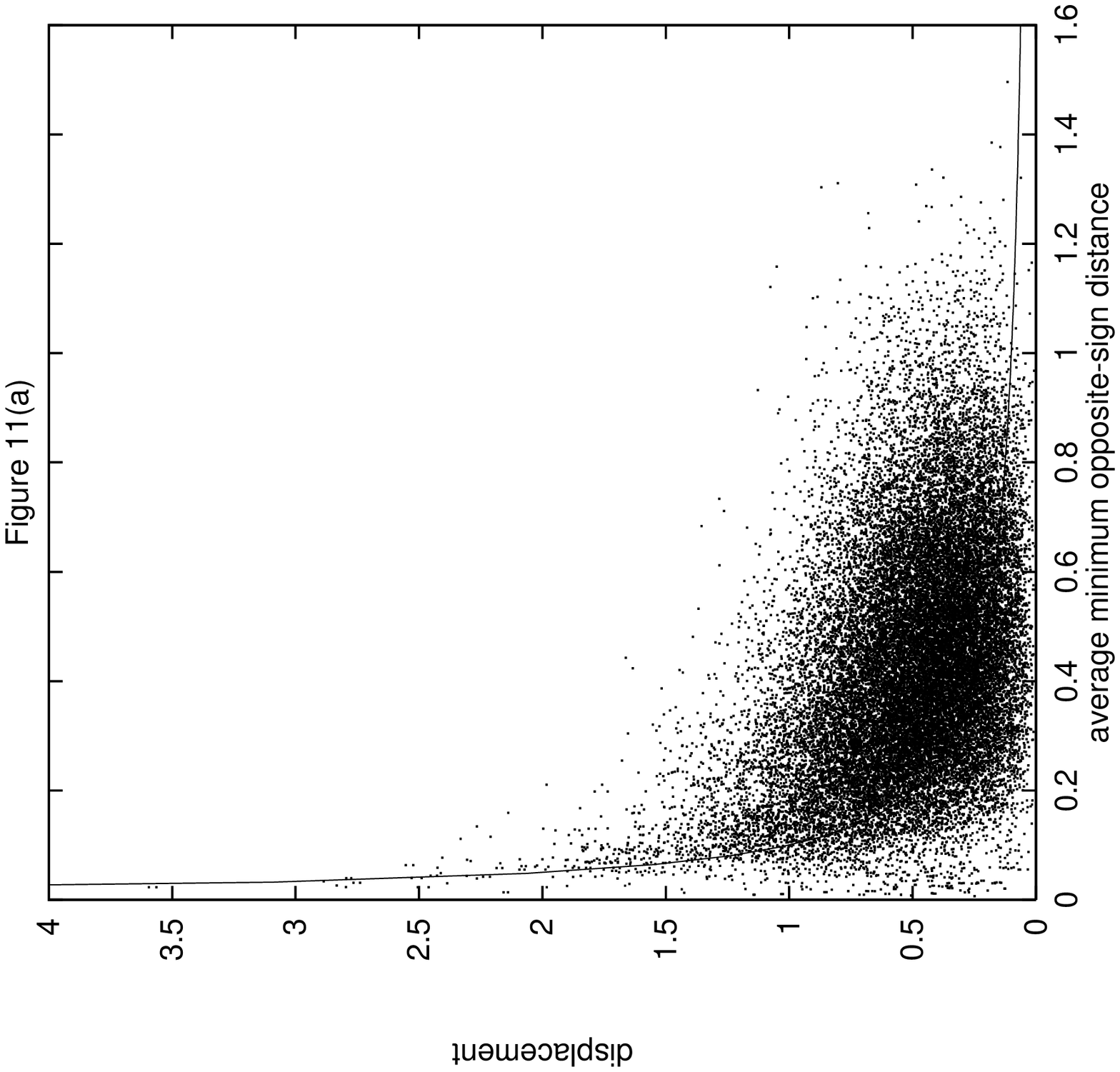}
\end{figure}

\begin{figure}
\epsfbox{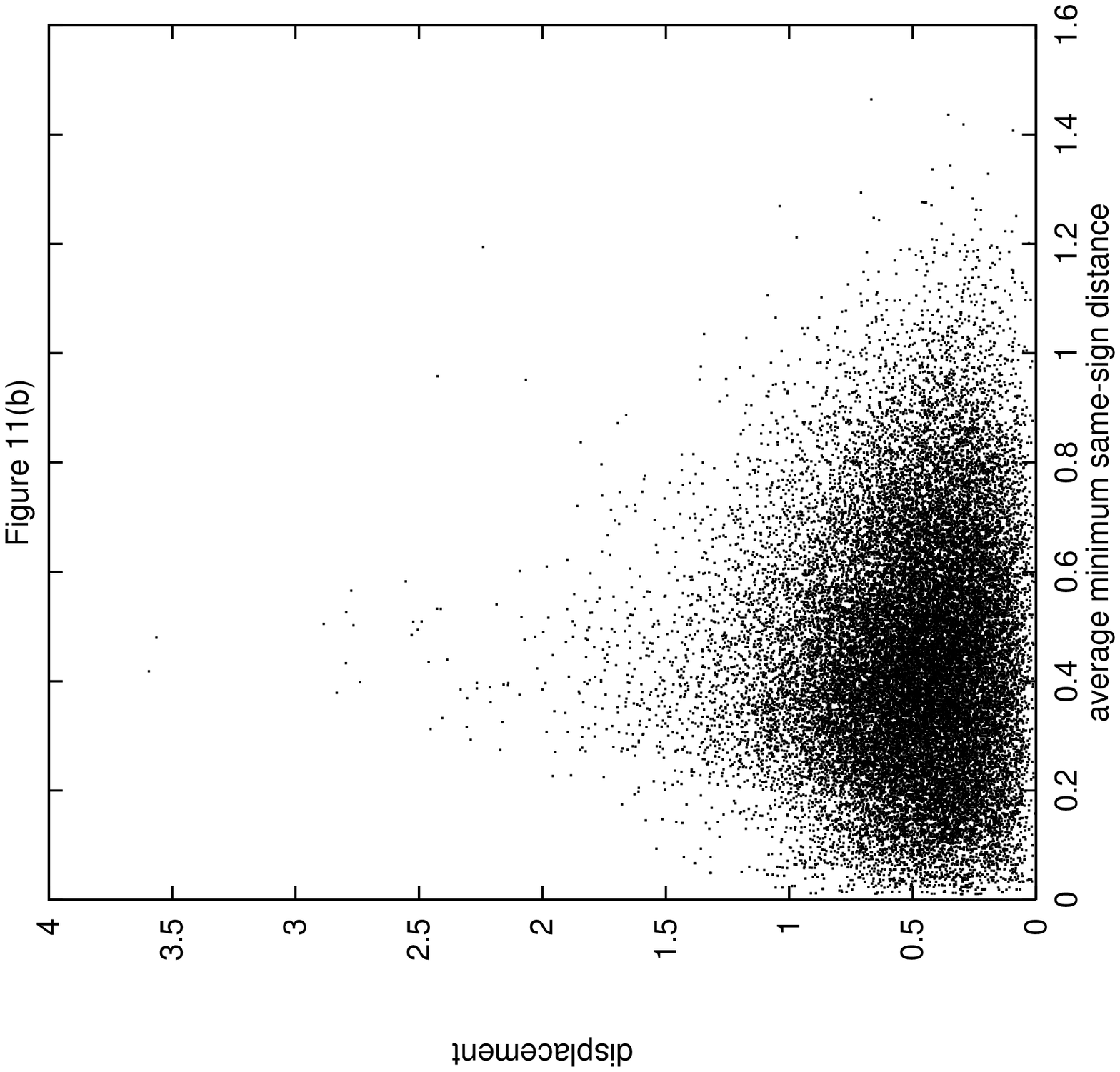}
\end{figure}

\begin{figure}
\epsfbox{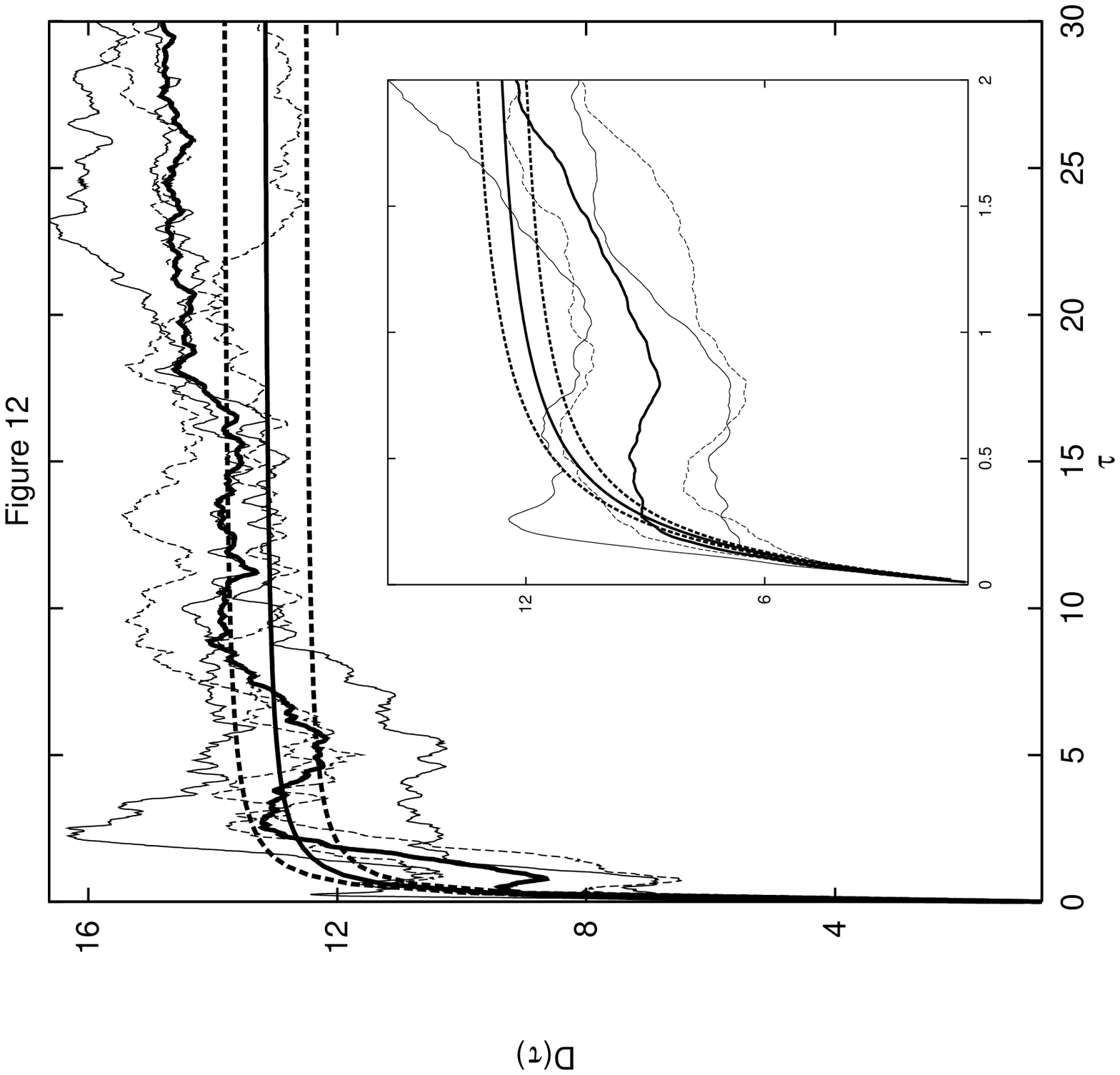}
\end{figure}

\end{document}